# Mechanical and Tribological Properties of Layered Materials Under High Pressure: Assessing the Importance of Many-Body Dispersion Effects

Wengen Ouyang,[1] Ido Azuri,[2] Davide Mandelli,[3] Alexandre Tkatchenko,[4,5] Leeor Kronik,[2] Michael Urbakh,[1] and Oded Hod[1]

[1] *School of Chemistry and The Sackler Center for Computational Molecular and Materials Science, Tel Aviv University, Tel Aviv 6997801, Israel*

[2] *Department of Materials and Interfaces, Weizmann Institute of Science, Rehovoth 76100, Israel*

[3] *Istituto Italiano di Tecnologia, Via Morego, 30 16163 Genova, Italy*

[4] *Fritz-Haber-Institut der Max-Planck-Gesellschaft, Faradayweg 4-6, D-14195 Berlin, Germany*

[5] *Department of Physics and Materials Science, University of Luxembourg, Luxembourg*

ABSTRACT

The importance of many-body dispersion effects in layered materials subjected to high external loads is evaluated. State-of-the-art many-body dispersion density functional theory calculations performed for graphite, hexagonal boron nitride, and their hetero-structures were used to fit the parameters of a classical registry-dependent interlayer potential. Using the latter, we performed extensive equilibrium molecular dynamics simulations and studied the mechanical response of homogeneous and heterogeneous bulk models under hydrostatic pressures up to 30 GPa. Comparison with experimental data demonstrates that the reliability of the many-body dispersion model extends deep into the sub-equilibrium regime. Friction simulations demonstrate the importance of many-body dispersion effects for the accurate description of the tribological properties of layered materials interfaces under high pressure.

Keywords: Van der Waals, density functional theory, many-body dispersion, interlayer potential, bulk modulus, hydrostatic pressure.

# 1. Introduction

Accurate modeling of the interlayer interactions in layered materials is of paramount importance for obtaining a quantitative description of their unique mechanical and tribological properties. Recently, a new generation of van der Waals (vdW) dispersion models was proposed within the framework of density functional theory (DFT) and their accuracy in predicting the binding energy and interlayer distance of two dimensional materials was assessed[1-4] against higher level methods such as the random phase approximation (RPA)[5, 6] and quantum Monte Carlo (QMC),[7] as well as against experiment.[8-11] Specifically, for layered materials that are not highly polarizable or ionic (e.g. graphene and hexagonal boron nitride ($h$-BN)), the Tkatchenko-Scheffler (TS) approach[12] and the many-body dispersion (MBD) method,[13, 14] in combination with the Heyd-Scuseria-Ernzerhof (HSE) hybrid density functional approximation,[15-17] were found to predict reliable equilibrium distances, binding energies,[1-3, 18] and elastic constants.[2, 3] So far, however, the accuracy of these vdW dispersion models has been tested mainly near the equilibrium configuration of model bilayers. In this work, we assess the accuracy of the TS and MBD vdW dispersion models at the equilibrium and sub-equilibrium interlayer distances regimes by combining state-of-the art DFT calculations with classical molecular dynamics (MD) simulations.

As test cases, which are relevant in view of many practical applications,[2, 19-24] we consider bulk graphite, bulk hexagonal boron nitride ($h$-BN), and their heterostructures. We first performed DFT calculations of binding energy (BE) curves and sliding potential energy surfaces (PES) for graphite, bulk $h$-BN, and their alternating heterostructures adopting two different methods: HSE+TS and HSE+MBD. From these two sets of reference data, we obtained two distinct parametrizations of our classical registry dependent interlayer potential (ILP),[25-28] which is able to accurately capture both BE curves and PES of these layered materials. Finally, we performed extensive equilibrium MD simulations under hydrostatic pressure ranging between 0 and 30 GPa, from which we extracted the interlayer distance as a function of the applied pressure ($c$-$P$ curve).

We find that the $c$-$P$ curves generated by the ILP parameters fitted against the HSE+MBD reference data compare well with experimental measurements,[29-35] for both graphite and bulk $h$-BN, over the entire range of pressures investigated. The bulk modulus extracted from the pressure-volume ($P$-$V$) curves also agrees well with experimental data. However, results obtained by adopting the ILP parameters fitted against the HSE+TS reference data deviate from experimental results, especially for graphite. Results of sliding friction simulations under high pressure further demonstrate the importance of an accurate description of the interlayer interactions in the sub-equilibrium regime for obtaining qualitatively and quantitatively correct results.

## 2. Methods

*2.1. DFT calculations*

We used the MBD and TS augmented HSE functional, as implemented in the FHI-AIMS code,[36] with the tier-2 basis-set,[37] using tight convergence settings, including all grid divisions and a denser outer grid. For the two dimensional (2D) systems, a vacuum of 50 Å was used with a *k*-grid of 19×19×1 points. For the MBD calculations, a large cutoff value of 1,300 Å was used for integrating the dipole field, as required for low-dimensional systems, together with a supercell cutoff of 45 Å. With these settings the MBD energy is converged to the level of $10^{-4}$ eV/atom. For the three dimensional (3D) systems, a *k*-grid of 19×19×7 points was used. The MBD convergence rate as a function of the cutoff parameters in the 3D calculations was faster compared to the 2D case. A smaller cutoff value of 300 Å and a supercell cutoff value of 30 Å were used. The MBD energy with this setting was estimated to be converged to $10^{-4}$ eV/atom, as well. At the high-pressure regime, the supercell cutoff radius had to be lowered from 30 Å to 25 Å in order to accelerate the calculations, however at that scale the effect of this reduction on the results was found to be negligible. In all cases, the HSE+TS energy was converged to $10^{-6}$ eV.

*2.2. Equilibrium MD simulation protocol*

To calculate the *c-P* curves of graphite and bulk *h*-BN, we adopted super-cell models consisting of twelve roughly square layers (5 nm×5 nm), each containing 880 carbon atoms or 440 boron + 440 nitrogen atoms, respectively. The layers in graphite are arranged in an alternating AB stacking, with a period *c* initially set equal to the experimental value of 6.70 Å.[30] The layers in bulk *h*-BN are arranged in an alternating AA' stacking (boron atop nitrogen), with a period *c* initially set equal to the experimental value of 6.66 Å.[32] Intra-layer interactions within each graphene and *h*-BN layer are modeled via the second generation REBO potential[38] and the Tersoff potential of Ref. [39], respectively. Interlayer interactions are modeled using the ILP or the Kolmogorov-Crespi (KC) potentials (for graphite), the construction of which is explained in details in Refs. [25-27], [40], [41], reparametrized herein to better describe the sub-equilibrium regime, as described below. All MD simulations were performed with the LAMMPS simulation package.[42] The velocity-Verlet integrator with a time-step of 1 fs was used to solve the equations of motion while enforcing periodic boundary conditions in all directions. A Nosé-Hoover thermostat with a time constant of 0.25 ps was used for constant temperature simulations. To maintain a specified hydrostatic pressure, the three translational vectors of the simulation cell were adjusted independently by a Nosé-Hoover barostat with a time constant of 1.0 ps.[43, 44] To generate the *c-P* curves we first equilibrated the systems in the NPT ensemble at a temperature of *T*=300 K and a fixed target pressure for 100 ps. After equilibration, the *c* lattice parameter was computed by averaging over a subsequent simulation period of 100 ps. The same procedure was repeated for different pressures ranging from 0 GPa to 30 GPa and the *c-P* curve was constructed. Tests with longer equilibration and averaging runs (200 ps + 200 ps) gave similar results.

*2.3. Definition of the interlayer distance for highly deformed surfaces*

For the alternating graphene/*h*-BN heterostructures, the out-of-plane deformation is large due to their intrinsic intralayer lattice vector mismatch. To calculate the *c-P* curves of this system, a new definition of the interlayer distance is required since the difference between the center-of-mass (COM) of the neighboring layers along *c*-axis is no longer a good measure. In the present study, to evaluate the interlayer distance for highly curved surfaces we first found for each atom, *i*, on a given layer its nearest neighbor, *j*, on the adjacent layer. Then we projected the vector connecting the pair along the local normal directions at the two atomic positions (see Ref. [40] for the definition of the normal vectors). The average between the two values is defined as the local distance between the layers. Further averaging over all positions *i* provides the value of the interlayer distance for a given configuration. At finite temperature, we also average over time to take into account thermal fluctuations. We note that for planar interfaces this definition matches the above-mentioned COM definition.

*2.4. Friction simulations*

To study the effects of external load on friction, we built 4-layer graphene and 4-layer *h*-BN homogeneous rectangular models with optimal stacking. The lateral dimensions of each model were 5 nm×5 nm and periodic boundary conditions were applied in both lateral directions. The rigid top layer (slider) was attached to a spring ($K_{dr} = 10$ N/m) moving at a constant velocity ($v_{dr} = 5$ m/s) along the zigzag direction and the bottom layer (substrate) was kept at rest. The force-fields used here were the same as those described above for the static calculations. A Langevin thermostat was added to the two internal layers and the damping coefficients used were $\eta_x = \eta_y = \eta_z = 1$ ps$^{-1}$. The systems were first equilibrated at 300 K for 400 ps with a time step of 1 fs, in absence of the pulling force, following which the friction simulations commenced. The static friction force is defined as the maximum force recorded across the entire force-trace and the kinetic friction force is calculated as $\langle F_{\text{kinetic}} \rangle = \langle K_{dr}(v_{dr}t - x_{\text{slider}}) \rangle$, where *t* is the simulation time, $x_{\text{slider}}$ is the position of center-of-mass of the slider along sliding direction and $\langle \cdot \rangle$ denotes a steady-state time average. The statistical errors have been estimated using 10 different datasets, each calculated over a time interval of 300 ps.

## 3. Force-Field Parameterization

The study of the properties of bulk graphite and *h*-BN under high pressure requires an interlayer potential (ILP) flexible enough to allow an accurate description of interactions in both equilibrium and sub-equilibrium regimes and, most important, to be able to describe the strongly anisotropic character of the layered materials under study. We chose our recently developed ILP,[27] for which we previously provided two sets of parameters for homogeneous and heterogeneous systems based on graphene and *h*-BN.[25, 26] We stress here that these sets of parameters were fitted manually against HSE+MBD reference data focusing on achieving good agreement only in the near-equilibrium and long-range interaction regime. More recently,[28] we have provided a refined set of parameters fitted using an automatic interior-point technique[45, 46] that allowed us to improve the agreement with the reference HSE+MBD data. Furthermore, in Ref. [28], we also provided a set of refined parameters for the KC potential[40] for graphene based systems. We note that all the above parametrizations have been benchmarked against DFT reference data calculated in a bilayer geometry, considering interlayer distances ranging from 2.5 Å to 15 Å.

*3.1. Binding energy curves*

Here, due to the importance of the sub-equilibrium interlayer distance regime for the tribological properties of layered materials, we perform new benchmark HSE+TS and HSE+MBD calculations for bulk graphite, *h*-BN, and their alternating heterostructures, which considering interlayer distances in the range of 2-10 Å with increased resolution. **Figure 1** presents binding energy curves calculated for the fully periodic structures of bulk graphite (first row), bulk *h*-BN (second row), and C-stacked alternating graphene/*h*-BN heterojunctions (third row) using HSE+MBD (left column, full black circles) and HSE+TS (right column, full black circles). The corresponding ILP fits are marked in red open squares. As may be expected, the HSE+TS approach provides deeper potential energy wells than HSE+MBD with similar equilibrium interlayer distances (see a detailed analysis in the Benchmark Tests section below). Notably, the ILP can be well fitted (using the procedure described in Ref. [28]) against both the pair-wise HSE+TS results and the many-body reference data throughout the entire interlayer distance range considered, which extends deep into the sub-equilibrium regime.

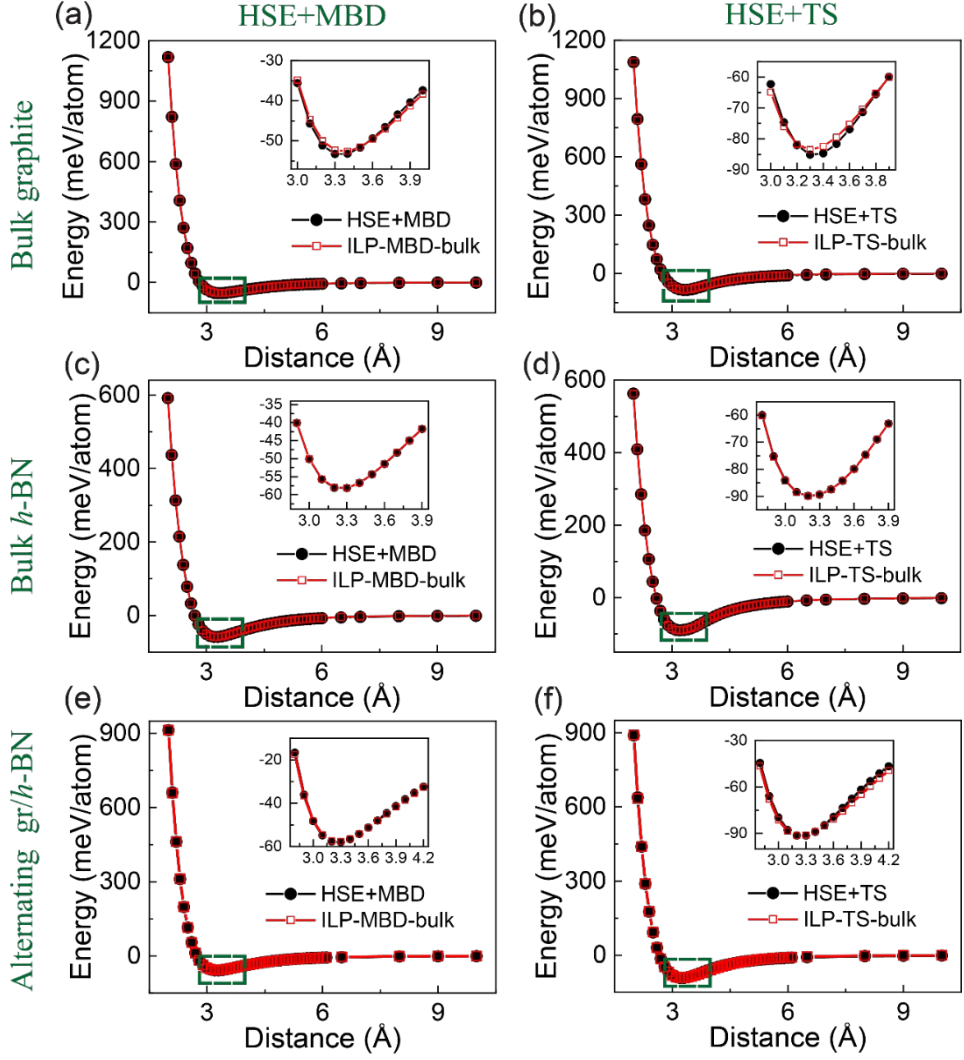

*Figure 1.* Binding energy curves of the fully periodic structures of bulk graphite (upper row), bulk h-BN (middle row), and C-stacked alternating graphene/h-BN heterojunctions (bottom row), calculated using HSE+MBD (left column, full black circles) and HSE+TS (right column, full black circles), along with the corresponding ILP fits (open red squares). The reported energies are measured relative to the value obtained for infinitely separated layers and are normalized by the total number of atoms in the unit-cell. The insets provide a zoom-in on the equilibrium interlayer distance region.

*3.2. Sliding potential energy surfaces*

The upper rows of **Figure 2** and **Figure 3** show the sliding PES of the three fully periodic structures considered, calculated at their equilibrium interlayer distances using HSE+MBD and HSE+TS, respectively. The corresponding ILP data appear in the middle row of both figures and the differences between the reference DFT data and the ILP results are presented in the lower panels. For all three systems, the HSE+MBD approach predicts somewhat lower PES corrugation than the HSE+TS method. The ILP fitting is in good qualitative and quantitative agreement with the DFT reference data.

Specifically, for the HSE+MBD results, the maximal deviation between the DFT reference and ILP results for bulk graphite is 4.7% of the overall PES corrugation. The corresponding differences for bulk *h*-BN and the heterogeneous structures are 0.25% and 4.2%, respectively.

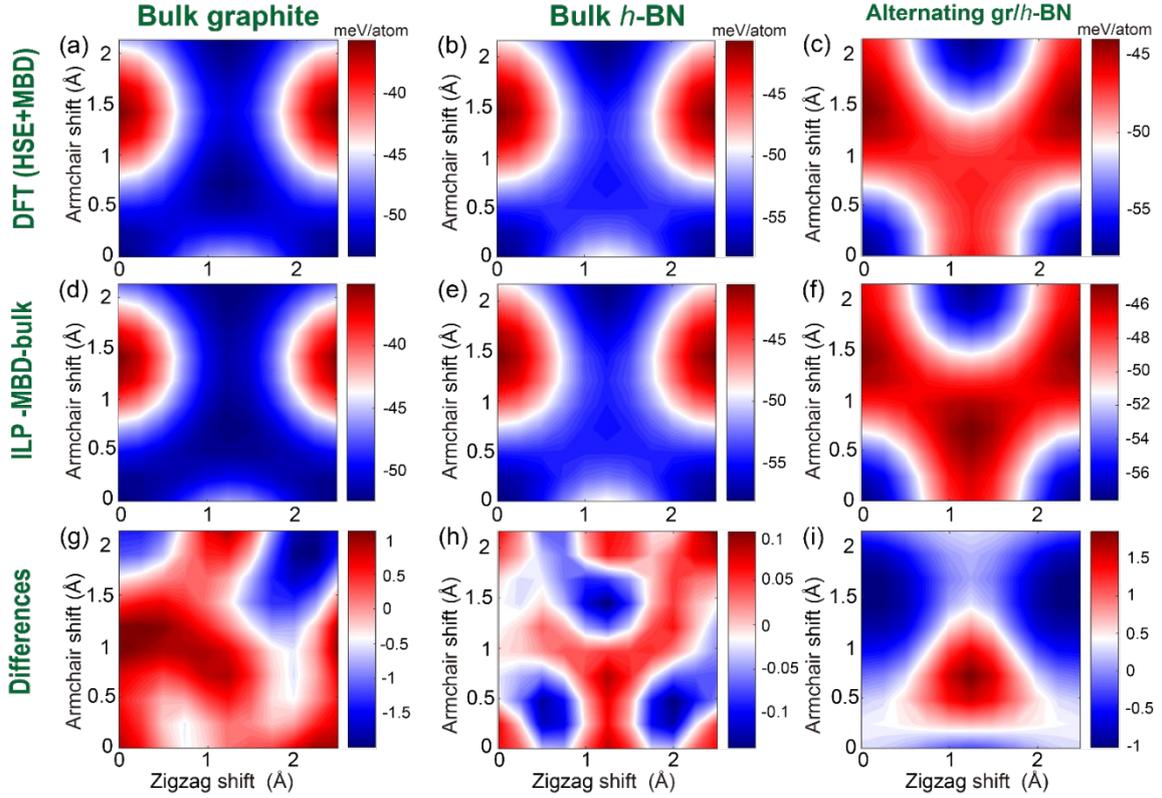

*Figure 2.* Sliding energy surfaces of the periodic structures considered, calculated at an interlayer distance of 3.3 Å. The first and second rows present the sliding energy surface of bulk graphite (left panels), bulk h-BN (middle panels), and alternating graphene/h-BN (right panels) systems, calculated using HSE+MBD and ILP-MBD-bulk parameterization, respectively. The third row presents their differences. The parameters of Table S1 in the SI are used for the ILP calculations. The reported energies are measured relative to value obtained for the infinitely separated layers and are normalized by the total number of atoms in the unit cell.

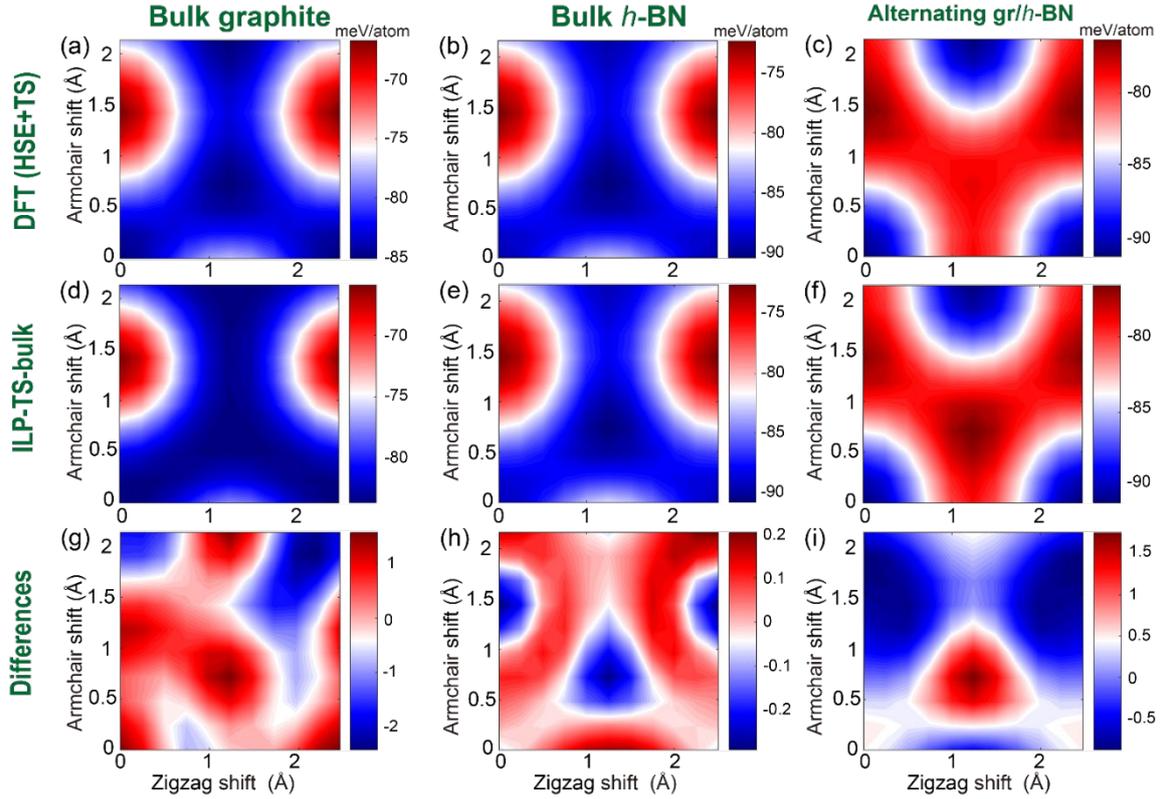

*Figure 3.* Sliding energy surfaces of the periodic structures considered, calculated at an interlayer distance of 3.3 Å. The first and second rows present the sliding energy surface of bulk graphite (left panels), bulk h-BN (middle panels), and alternating graphene/h-BN (right panels) systems, calculated using HSE+TS and ILP-TS-bulk parameterization, respectively. The third row presents their differences. The parameters of Table S2 in the SI are used for the ILP calculations. The reported energies are measured relative to the value obtained for the infinitely separated layers and are normalized by the total number of atoms in the unit cell.

Notably, while the ILP with its present parameterizations captures well all the main symmetries of the full sliding energy surface corrugation, it cannot generally be expected to capture the symmetry of the dispersive component alone (see Section 5 of the Supplementary information (SI)). This is because the sliding energy corrugation associated with this component is found to be typically lower than 2 meV/atom, which is below the expected accuracy of the ILP for these systems (see lower panels of **Figure 2** and **Figure 3**). In cases where Pauli repulsions dominate the sliding energy surface, such as those used as reference for the present ILP parameterization, this has a negligible effect. However, in scenarios where the sliding energy corrugation is dominated by the dispersive component this may have an important effect. Such scenarios can be encountered in large moiré superstructures characterized by high surface undulations that result in large interlayer separations, which may require a dedicated parameterization of the ILP.

*3.3. Parameters*

All fitting parameters can be found in Section 1 of the SI. We mark the new ILP parameterizations presented herein as ILP-TS-bulk and ILP-MBD-bulk. For clarity, we name the original ILP parameters of Refs. 25, 26, and the refined parameters of Ref. 28, both fitted against bilayer calculations, as ILP-MBD-bilayer-original and ILP-MBD-bilayer-refined, respectively. Finally, we name the original parameters of the KC potential of Ref. 40 and the refined ones of Ref. 28 as KC-original and KC-MBD-bilayer-refined, respectively. The sensitivity test of the ILP parameters is provided in Section 2 of the SI.

## 4. Benchmark tests

*4.1. Compressibility*

The simulation results for graphite and bulk *h*-BN under hydrostatic pressure are presented in **Figure 4**, along with the experimental *c-P* curves. We note that the slope of the normalized *c-P* curve for graphite, akin to the compressibility $\sim \partial[(c(P) - c_0)/c_0]/\partial P$, predicted by the ILP-TS-bulk parametrization (green stars in panel (a)) deviates from the experimental one at loads $\gtrsim 4$ GPa, systematically overestimating the experimental values obtained under hydrostatic pressure (up and down oriented violet triangles, cyan pentagons, black circles). A similar behavior is observed also for the case of *h*-BN. The *c-P* curves obtained from the ILP-MBD-bulk parametrization (open diamonds in panels (a) and (b)) somewhat deviate from the experimental data only at considerably high pressures, $\gtrsim 20$ GPa and $\gtrsim 8$ GPa, respectively, for graphite and *h*-BN. Notably, the ILP parameterizations performed against bilayer (ILP-MBD-bilayer-refined, open circles in panel (a)) and bulk graphite (open diamonds) reference data provide comparably good agreement with the experimental data, indicating that interactions between non-adjacent graphene layers are negligible. Somewhat larger differences are observed for the case of *h*-BN (see open circles and open diamonds in panel (b)). We further observe that if the ILP is not well parameterized in the sub-equilibrium regime, the obtained *c-P* curves deviate significantly from the experimental data. See, for example, the deviation of the ILP-MBD-bilayer-original results for bulk *h*-BN (open brown squares) from the experimental values, in the right panel of **Figure 4**. We note that the ILP-MBD-bilayer-original parameterization provides a relatively good fit for bilayer graphene down to 2.8 Å and thus the c-P curves calculated for graphite are in good agreement with the experimental data in the presented pressure range. Finally, while the original KC interlayer potential for graphite (brown triangles in panel (a)) loses accuracy at high pressures ($\gtrsim 4$ GPa), our new KC potential parameterization (blue triangles) yields results in agreement with the ILP-MBD-bulk parametrization.

The main conclusion that can be drawn from the above results is that the new MBD ILP parametrization, which extends down to an interlayer distance of 2 Å, performs very well across the entire pressure range investigated. This indicates that the DFT reference data are reliable even in the deep sub-equilibrium regime, where bulk graphite and *h*-BN are compressed down to 0.6 of their equilibrium interlayer distance.

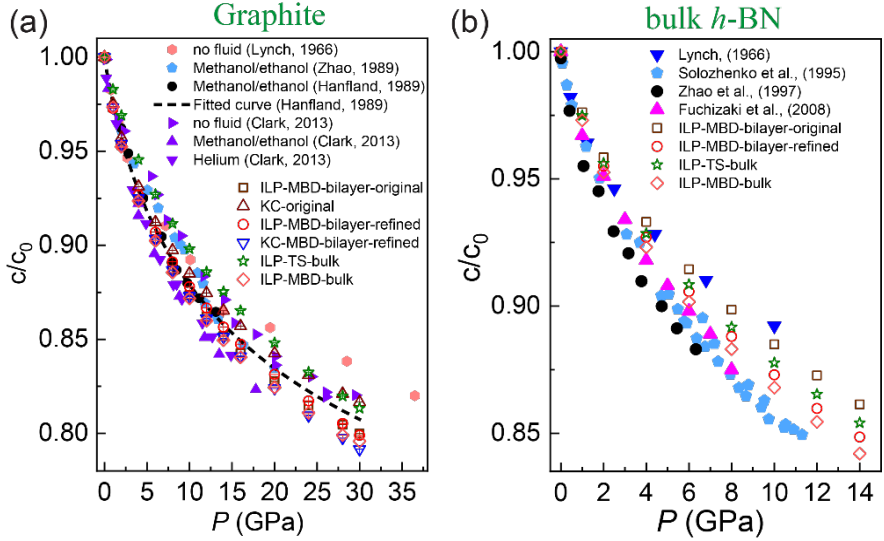

*Figure 4. Measured and computed pressure dependence of the c lattice parameter of (a) bulk graphite and (b) bulk h-BN. Each result is normalized by the zero pressure value, $c_0$, corresponding to the same measurement or computation. Full symbols represent experimental results and open points represent NPT simulation results for different parameterizations of the ILP and KC potentials, as specified in the corresponding set labels. Error bars for the simulated data, obtained from the temporal standard deviation of the interlayer distance thermal fluctuations at equilibrium, are smaller than the symbol width.*

### 4.2. Bulk moduli

To verify that the HSE+MBD ILP parametrization, including the high-pressure regime, does not harm its ability to predict low-pressure bulk properties, we calculate the bulk moduli of bulk graphite and *h*-BN and compare against experimental values. The computed bulk moduli are obtained by fitting our simulations data across the entire pressure range considered to the Murnaghan equation of state (EOS):[30, 47]

$$V(P)/V_0 = [1 + (B'_V/B^0_V)P]^{-1/B'_V}. \tag{1}$$

Here, $V_0$ and $V(P)$ are the unit-cell volumes in the absence and presence of an external hydrostatic pressure, $P$, and $B^0_V$ and $B'_V$ are the bulk modulus and its pressure derivative at zero pressure, respectively. The corresponding Murnaghan fits for the various ILP and KC parameterization results can be found in Section 3 of the SI. **Figure 5** shows experimental $V(P)$ curves, along with those obtained by the various ILP parameterizations considered above and the corresponding fits of the ILP-TS-bulk and ILP-MBD-bulk results to Eq. (1). While high-pressure experimental volumetric data is less abundant than inter-plane lattice constants information, especially for graphite, we find that the MBD parameterized ILP results are overall in better agreement with the most recent experimental data, across the pressure range considered. The extracted bulk moduli for bulk graphite and *h*-BN, along with their zero pressure derivatives, binding energies, and lattice constants are compared in Table 1.

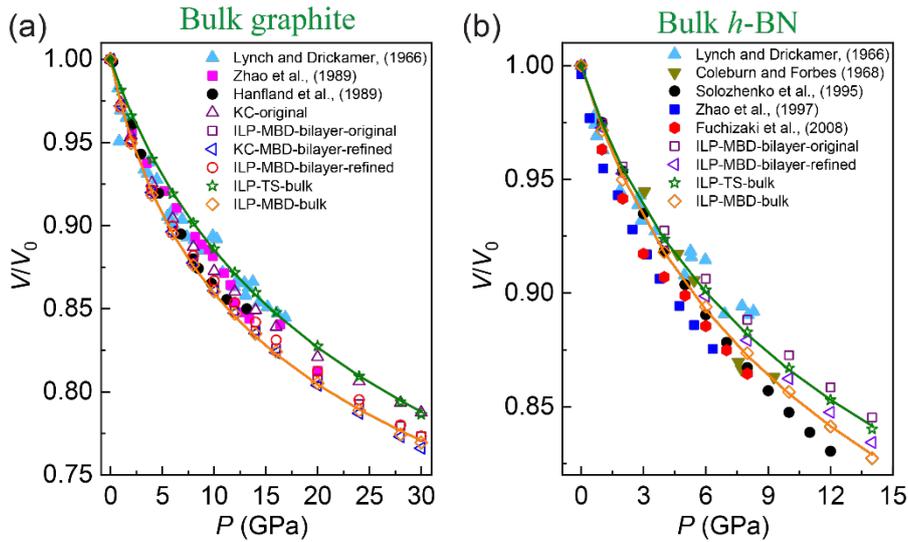

*Figure 5.* *Murnaghan plot for bulk graphite and bulk h-BN. Full points denote experimental results with different measurement methods and open symbols denote NPT simulations results for different parameterizations of the ILP and KC potentials, as specified in the corresponding set labels. For clarity of presentation we also show Murnaghan fitting curves of our simulation results obtained with the ILP-MBD-bulk (orange line) and ILP-TS-bulk (olive line) parameterizations.*

As can be seen in Table 1, the values of bulk modulus predicted for graphite by the MD simulations, using any of the MBD ILP parametrizations agree well with the experimental data (ranging between 30.8 GPa and 33.8 GPa). They are also in good agreement with previous PBE+MBD@rsSCS calculations, predicting a bulk modulus of 29 GPa.[3] Furthermore, both the original, empirically fitted KC potential and our MBD parameterization of it provide reasonable agreement with a slight overestimation of the experimental bulk modulus. Comparing the MBD results to other first-principles methods we obtain the commonly observed behavior, where the local density approximation (LDA) provides reasonable prediction of the bulk modulus of graphite (but not necessarily for the right reasons) while the Perdew-Burke-Ernzerhof (PBE) gradient-corrected exchange correlation density functional approximation strongly underestimates it. RPA calculations overestimate the experimental values by ~10%, the empirical Grimme D2 pair-wise dispersion correction overestimates the modulus by ~15%, and all TS-based calculations (including our TS parameterized ILP) overestimate the bulk modulus of graphite by up to 80%. Considering the bulk modulus derivative with respect to the external pressure, we find that all MBD force-field parameterizations, as well as the empirically fitted KC potential, provide good agreement with the lower experimental value of 8.9±0.1. The only outlier within our test-set is the TS parameterized ILP, which underestimates the bulk modulus pressure derivative by nearly 30%.

Table 1. Bulk modulus $(B_V^0)$ and its zero pressure derivative $(B_V')$, intra- $(a_0)$ and inter- $(c_0)$ layer lattice constants and binding energy $(E_{bind})$ of bulk graphite, calculated using the various force-field parameterizations and compared to experimental and first-principles values.

| | Methods | $B_V^0$ (GPa) | $B_V'$ | $a_0$ (Å) | $c_0$ (Å) | $E_{bind}$ (meV/atom) |
|---|---|---|---|---|---|---|
| Experiments | X-ray diffraction, Ref. 29 | 32±2[a] | 12.3±0.7[a] | 2.4612 | 6.7078 | -- |
| | X-ray diffraction, Ref. 30 | 33.8±0.3 | 8.9±0.1 | 2.603(4) | 6.706(3) | -- |
| | X-ray diffraction, Ref. 31 | 30.8±2 | -- | 2.462 | 6.707 | -- |
| First-principles | LDA+PAW/US, Ref. 48 | 30.30/28.98 | -- | 2.46 | 6.78 | -- |
| | LDA, LCGTO-FF, Ref. 49 | 50.9[b], 38.3[c] | -- | -- | -- | -- |
| | PBE, Ref. 50 | 1 | -- | 2.47 | 8.84 | 1 |
| | PBE+D2, Ref. 50 | 38 | -- | 2.46 | 6.45 | 55 |
| | RPA, Ref. 51 | 36 | -- | -- | 6.68 | 48 |
| | QMC, Ref. 7 | -- | -- | 2.4595 | 6.85(7) | 56(5) |
| | PBE+TS, Ref. 52 | 56 | -- | 2.46 | 6.65 | -- |
| | PBE+TS, Ref. 53 | 59 | -- | 2.46 | 6.68 | 82 |
| | PBE+TS/SCS, Ref. 53 | 43 | -- | 2.46 | 6.75 | 55 |
| | PBE+TS/HI, Ref. 1 | 57 | -- | 2.46 | 6.74 | 81 |
| | PBE+MBD@rsSCS, Ref. 3 | 29 | -- | 2.46 | 6.82 | 48 |
| | HSE+TS, This paper | -- | -- | 2.462 | 6.60 | 85.12 |
| | HSE+MBD, This paper | -- | -- | 2.462 | 6.60 | 53.29 |
| MD Simulations[d] | ILP-MBD-bulk | 34±2 | 8.1±0.3 | 2.46031(2) | 6.8036(6) | 51.51(1) |
| | ILP-TS-bulk | 55±2 | 6.2±0.3 | 2.45934(4) | 6.6605(7) | 81.97(2) |
| | ILP-MBD-bilayer-refined, Ref. 28 | 34±3 | 8.1±0.6 | 2.46027(5) | 6.742(2) | 50.73(3) |
| | ILP-MBD-bilayer-original, Ref. 25 | 33±1 | 8.5±0.3 | 2.46026(4) | 6.768(1) | 50.58(2) |
| | KC-MBD-bilayer-refined, Ref. 28 | 35±2 | 7.7±0.4 | 2.46029(3) | 6.788(1) | 51.21(3) |
| | Original KC, Ref. 40 | 37±2 | 8.9±0.4 | 2.46036(5) | 6.752(2) | 46.18(3) |

[a]Fit with Eq. (1); [b]EOS fit; [c]cubic fit; [d]The MD simulations were performed at 300 K.

All DFT and force-field parameterizations appearing in Table 1 provide good agreement with the experimental values of the intra- and interlayer lattice constants. The accuracy of both first-principles and force-field predictions of the intralayer lattice constant is found to be ~0.01 Å, whereas the accuracy of the interlayer lattice constant is within ~0.1 Å, apart from PBE and PBE+D2 that overestimate and underestimate the interlayer lattice constant, respectively. Finally, all MBD calculations and force-field parameterizations provide bulk graphite binding energies within 10% of both RPA and QMC results. Nonetheless, all TS calculations overestimate the binding energy by nearly 50%.

Table 2. Bulk modulus $(B_V^0)$ and its zero pressure derivative $(B_V')$, intra- $(a_0)$ and inter- $(c_0)$ layer lattice constants and binding energy $(E_{bind})$ of bulk h-BN, calculated using the various force-field parameterizations and compared to experimental and first-principles values.

| | Methods | $B_V^0$ (GPa) | $B_V'$ | $a_0$ (Å) | $c_0$ (Å) | $E_{bind}$ (meV/atom) |
|---|---|---|---|---|---|---|
| Experiments | X-ray diffraction, Ref. 54 | -- | -- | 2.50399(5) | 6.6612(5) | -- |
| | X-ray diffraction, Ref. 29 | 22±4[a] | 18±3[a] | 2.5040 | 6.6612 | -- |
| | X-ray diffraction, Ref. 32 | 36.7±0.5 | 5.6±0.2 | 2.504(2) | 6.660(8) | -- |
| | X-ray diffraction, Ref. 33 | 17.6±0.8 | 19.5±3.4 | 2.5043(1) | 6.6566(6) | -- |
| | X-ray diffraction, Ref. 55 | 27.6±0.5 | 10.5±0.5 | 2.504(4) | 6.659(2) | -- |
| | X-ray scattering, Ref. 56 | 25.6±0.8 | -- | 2.506 | 6.657 | -- |
| | X-ray diffraction, Ref. 34 | 21 | 16 | 2.50(5) | 6.66(3) | -- |
| First-principles | Theory, 57 | 27.7±0.2 | 9.0±0.1 | -- | -- | -- |
| | LDA, Ref. 58 | 30.1 | 10.1 | 2.496 | 6.4896 | 57 |
| | RPA, Ref. 59 | -- | -- | -- | 6.60 | 39 |
| | PBE+D2, Ref. 50 | 56 | -- | 2.51 | 6.17 | 77 |
| | PBE+TS, Ref. 52 | 37 | -- | 2.51 | 6.71 | -- |
| | PBE+TS, Ref. 53 | 36 | -- | 2.50 | 6.64 | 87 |
| | PBE+TS/SCS, Ref. 53 | 34 | -- | 2.50 | 6.67 | 73 |
| | PBE+TS/HI, Ref. 1 | 23 | -- | 2.51 | 6.78 | 62 |
| | PBE+MBD@rsSCS, Ref. 3 | 30 | -- | 2.50 | 6.59 | 59 |
| | HSE+TS, This paper | -- | -- | 2.500 | 6.40 | 89.85 |
| | HSE+MBD, This paper | -- | -- | 2.500 | 6.60 | 58.17 |
| MD Simulation[b] | ILP-MBD-bulk | 33±2 | 7.8±0.6 | 2.4959(1) | 6.6035(2) | 56.33(3) |
| | ILP-TS-bulk | 35±2 | 8.7±0.6 | 2.49334(6) | 6.5111(7) | 88.24(2) |
| | ILP-MBD-bilayer-refined, Ref. 28 | 35±2 | 8.0±0.6 | 2.49513(5) | 6.5461(2) | 57.38(3) |
| | ILP-MBD-bilayer-original, Ref. 25 | 38±3 | 8.7±0.9 | 2.49561(5) | 6.5817(7) | 56.32(1) |

[a]Fit with Eq. (1); [b]The MD simulations were performed at 300 K.

The experimental values of the bulk modulus and its pressure derivative for bulk h-BN are more scattered than those for graphite, ranging from 17.6 GPa to 36.7 GPa and 5.6 to 19.5, respectively. Therefore, it is difficult to draw a definite conclusion regarding the method that provides best results. Nevertheless, all methods listed in Table 2, apart from PBE+D2, yield values within the experimentally measured range.

Similar to the case of graphite, all DFT and force-field parameterizations appearing in Table 2 provide good agreement with the experimental values of the intra- and interlayer lattice constants. The accuracy of both first-principles and force-field predictions of the intra-layer lattice constant is found to be ~0.01 Å, whereas the accuracy of the interlayer lattice constant is within ~0.1 Å, apart from the PBE+D2 value that underestimate the interlayer lattice constant by ~0.5 Å. Finally, all MBD calculations and force-field parameterizations provide bulk $h$-BN binding energies that are ~ 44% above the RPA results and the corresponding TS calculations overestimate the binding energy by more than a factor of 2. We note, however, that these deviations may result in part from the approximate nature of the RPA calculation itself. We further note that the experimental values listed in Tables 1-2 were obtained by adopting different approximations for the EOS (see section 3 of SI for details). In Table S3 of section 3 of the SI we provide the elastic moduli obtained by fitting our $P$-$V$ curves using three different versions of the EOS. We found that all the EOS yielded consistent values of the bulk modulus. This suggests that the differences between the various experimental values of the bulk modulus arise from the different methods adopted to collect the data, rather than from the choice of the EOS used for their fitting, in contrast with the observation reported in Ref. 60.

Overall, we find that even when parameterized against extremely high pressure HSE+MBD reference data, the ILP provides good agreement with the experimental data for all bulk parameters considered. The fact that the corresponding TS-parameterized ILP fails to predict several bulk parameters indicates the importance of including MBD effects in the calculation and validates the reliability of the HSE+MBD method for describing graphitic and $h$-BN-based systems at both low and high external pressures.

*4.3. Phonon spectra*

To further demonstrate the ability of the newly parameterized HSE+MBD ILP to predict low-pressure properties, we computed the phonon dispersion curves of graphite and bulk $h$-BN at zero pressure and temperature, and compared them with the experimental data reported in Refs. 61 and 62, respectively. The results reported in **Figure 6**a-b show that the dispersion of the low energy out-of-plane (ZA) branches, which are related to the soft flexural modes of the layers, is well described for both bulk graphite and bulk $h$-BN (see **Figure 6**c-d). The larger deviations from the experimental data, observed for the high energy transverse (TO) and longitudinal (LO) optical modes, are mainly caused by the intralayer potential used in our simulations. More details can be found in Ref. 63. In contrast, while the HSE+TS parameterized ILP provides a good description of the low-frequency phonon spectrum of bulk $h$-BN, large deviations from the low energy experimental ZA branches are obtained for bulk graphite.

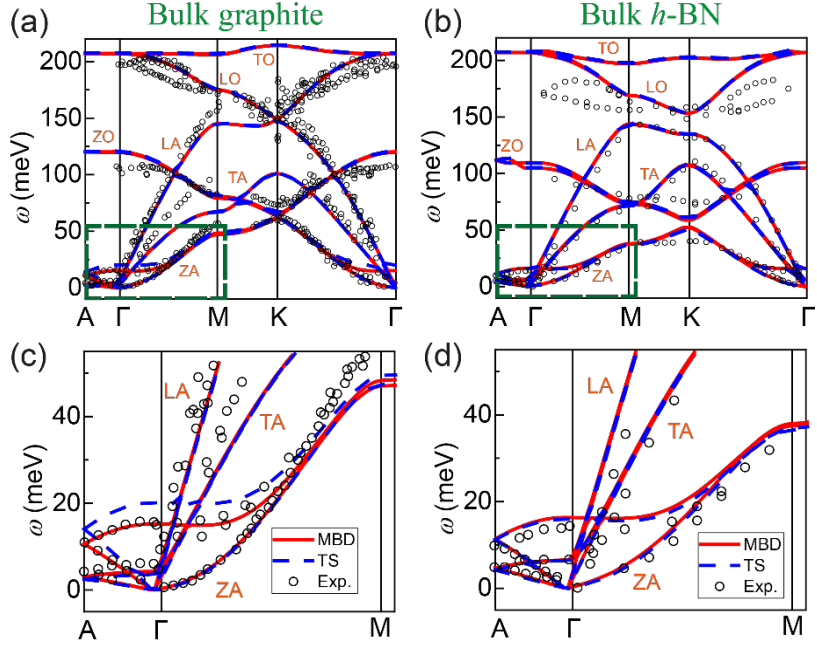

**Figure 6.** (a) Phonon spectra of (a) bulk graphite and (b) bulk h-BN. Red solid lines and blue dashed lines are dispersion curves calculated using the ILP with parameters listed in Table S1 (ILP-MBD-bulk) and Table S2 (ILP-TS-bulk) in the SI, respectively. Experimental results of bulk graphite[61] and bulk h-BN[62] are given by open black circles. Panels (c) and (d) show a zoom-in of the low energy phonon modes around the Γ-point (green rectangles in panels (a) and (b)) for graphite and h-BN, respectively.

## 5. Applications

*5.1. Heterogeneous graphene/h-BN stacks under high pressure*

In the previous sections, we have analyzed the performance of MBD based ILP parameterizations for predicting the mechanical properties of homogenous graphene and *h*-BN based structures under high-pressure. Using the same protocol, here we predict the behavior of two bulk heterogeneous structures formed between graphene and *h*-BN. The first one consists of twelve alternating layers of graphene and *h*-BN with aligned lattice vectors. We will refer to this model as 6-G/BN (see **Figure 7**a). The second model is constructed by stacking a six-layer graphite slab with AB stacking atop a six-layer *h*-BN slab with AA' stacking, in an aligned configuration. We name this model 6-G/6-BN (see **Figure 7**b). For both systems periodic boundary conditions are applied in all three directions. Due to their in-plane lattice mismatch of ~1.8%, graphene and *h*-BN form an incommensurate interface. In order to satisfy lateral periodic boundary conditions while preserving the experimental lattice mismatch, we followed the method outlined in Ref. 24 and built large rectangular supercells, where each graphene and *h*-BN layer contains 12,544 and 12,120 atoms, respectively. For both heterojunctions, we performed simulations using the ILP-MBD-bulk parameterization.

**Figure** 7a-b reports snapshots of the 6-G/BN and 6-G/6-BN models at 300 K and zero pressure. In contrast to the nearly flat (maximal corrugation of ~0.8 Å) homogeneous junctions, the heterogeneous systems exhibit large out-of-plane deformations. In particular, the alternating stack exhibits vertical distortions of the order of ~10 Å and the 6-G/6-BN model displays deformations of ~3 Å. These large deformations result from the delicate interplay between the intra-layer elastic energy contribution and the long-range interlayer dispersion interactions within the incommensurate junction.[26, 64] The difference in the out-of-plane deformations of the two heterojunctions results from the fact that the bending rigidity of a 6-layer stack of graphene or $h$-BN is higher than the bending rigidity of the individual layers.

**Figure** 7c shows the pressure dependence of the average graphene/$h$-BN interlayer spacing of the two heterojunctions (see Methods section for our definition of the interlayer spacing in corrugated structures). The corresponding results for bulk graphite and bulk $h$-BN are also plotted for comparison purposes. We find that the average interlayer distances of both heterostructures are consistently larger than that of bulk $h$-BN and similar to that of graphite. Furthermore, on average the interlayer distance of the 6-G/6-BN junctions is slightly larger than that of the 6-G/BN system.

Beyond thermal fluctuations, the inherently corrugated heterostructures exhibit a distribution of interlayer distances, the error bars in panel (c) illustrate the standard deviation of the distribution at zero pressure, which remains nearly constant for finite pressures (not shown for clarity). Notably, for both heterostructures, these distributions are considerably wider than the thermally broadened interlayer distance distributions of the homogeneous counterparts (see **Figure** 7d). The bulk modulus and its pressure derivatives obtained by fitting the results to the Murnaghan $V(P)$ equation produce similar values for both materials (see Table 3), which are comparable also to those of the corresponding homogeneous bulk structures.

As illustrated in panels (e) and (f) in **Figure 1**, the zero temperature equilibrium interlayer distances and binding energies predicted by HSE+MBD for the aligned bulk alternating C-stacked graphene/$h$-BN heterojunction are comparable to those of the homogeneous bulk values, giving 3.3 Å and 58.0 meV/atom, respectively. As may be expected, due to thermal fluctuations the corresponding room temperature average interlayer distance is somewhat larger (3.348 ± 0.004 Å, with the small uncertainty reflecting the negligible effect of thermal fluctuations in this case) and the binding energy is lower (52.59±0.04 meV/atom). Similarly, for the 6-G/6-BN heterostructure, the room-temperature MBD simulations predicted equilibrium interlayer distance and binding energy of 3.386 Å±0.005 Å and 53.28±0.02 meV/atom, respectively.

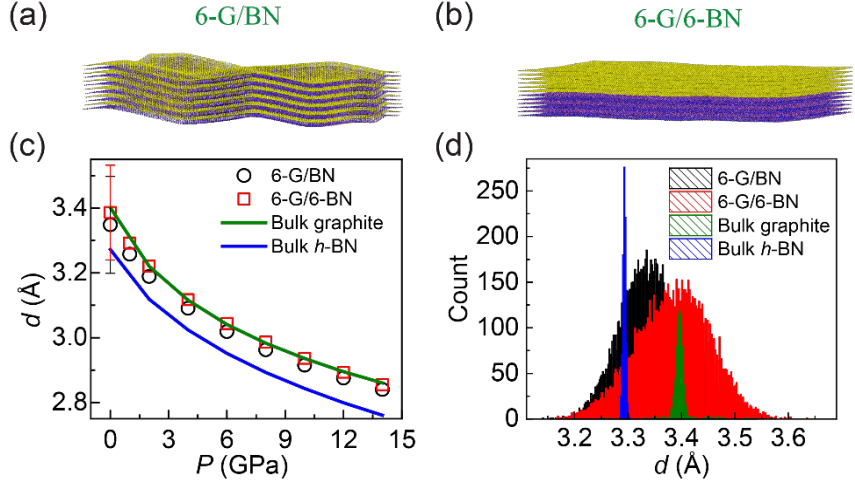

*Figure 7.* Snapshots of the (a) 6-G/BN and (b) 6-G/6-BN systems at zero pressure and 300 K. Individual graphene and h-BN layers are colored in yellow and blue, respectively. (c) Pressure dependence of the interlayer spacing, d, of 6-G/BN and 6-G/6-BN. ILP-MBD-bulk results for the homogeneous bulk graphene (full green line) and bulk h-BN (full blue line) systems are presented for comparison. (d) Distribution of the interlayer distance (averaged over time) of 6-G/BN (black) and 6-G/6-BN (red) at 300 K and zero external pressure. Corresponding results for the homogeneous bulk graphite (green) and bulk h-BN (blue) are presented for comparison. The standard deviation of the distribution in panel (d) defines the error bars in panel (c).

Table 3. Bulk modulus of heterogeneous structures calculated using the ILP-MBD-bulk parameterization.

| Structure | $B_V^0$ (GPa) | $B_V'$ |
|---|---|---|
| **6-G/BN** | 31±1 | 9.0±0.4 |
| **6-G/6-BN** | 32±1 | 8.4±0.4 |

*5.2. Normal load dependence of friction*

To evaluate the effects of the accuracy of the ILP in the sub-equilibrium regime for a practical dynamical application, we calculated the normal load dependence of friction in homogeneous graphene and *h*-BN sliding interfaces. We adopted three sets of ILP parameterizations: the original bilayer parameterization of Refs. 25, 26, the refined bilayer parameterization of Ref. 28, and the MBD bulk parameterization presented herein. Details of the simulation setup are given in the Methods section.

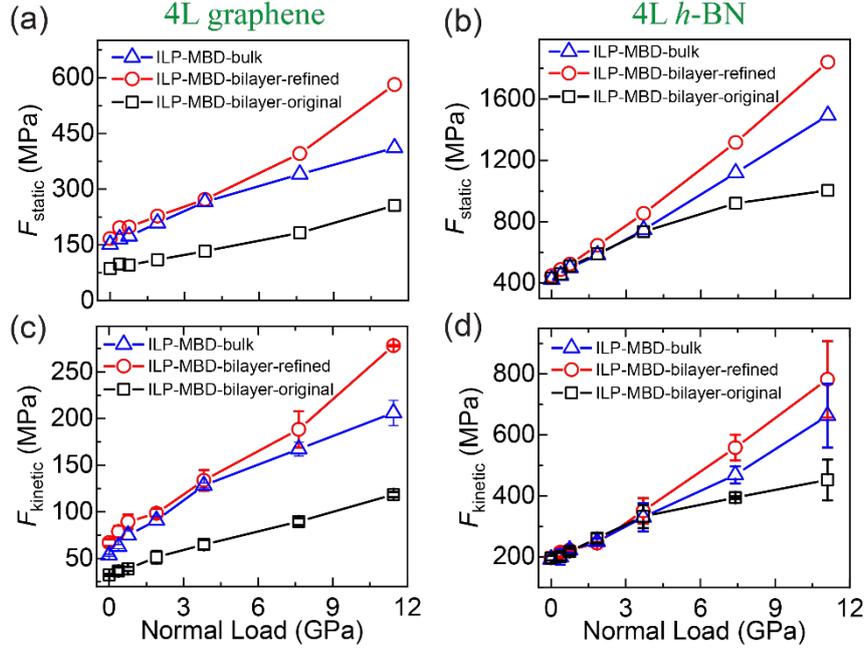

***Figure 8.*** *Normal load dependence of the static friction (top row) and kinetic friction (bottom row) for bulk graphite (left column) and bulk h-BN (right column). The simulations are performed at a temperature of 300 K using the MBD-bulk (open blue triangles), MBD-bilayer-refined (open red circles), and MBD-bilayer-original (open black squares) ILP parameterizations. See subsection 2.4 of the methods section for the error evaluation procedure.*

As can be clearly seen in **Figure 8**, for the four-layer graphene model the original bilayer parameterization predicts consistently lower static and kinetic friction-forces compared to the refined bilayer parameterization, across the entire load range considered with increasing deviations at the higher loads regime. This results from the fact that the two parametrizations provide similar agreement with the reference DFT binding data near the equilibrium interlayer distance but deviate at the sub-equilibrium regime. There, the automatic fitting procedure utilized in the refined parameterization provides better agreement with the reference data. As the same automatic fitting procedure is utilized also in the new MBD bulk parameterization it is found to be in good agreement with the refined bilayer parameterization results at the lower pressure regime. With increasing pressure, the overall interlayer distance decreases and next-nearest neighboring layers interactions in the bulk configuration become more important. This is reflected by the fact that in this regime, the bulk parameterization provides somewhat lower friction force values than the refined bilayer ILP. A very similar behavior is found for the four-layer *h*-BN system, but with better agreement between the three ILP parameterization up to an external pressure of ~3 GPa. This is consistent with the differences between the sliding PES and sliding energy barriers obtained by the various parameterizations for the studied junctions (see sections 4 and 6 of the SI).

Notably, for the case of *h*-BN, the original ILP parametrization predicts a sub-linear variation of the friction forces with pressure, whereas the new parameterization exhibits a linear behavior. This difference in qualitative behavior of the frictional properties stands in contrast with the compressibility results presented above, which were found to be less sensitive to the choice of ILP parameterization. This, in turn, further emphasizes the importance of an accurate description of the interlayer interactions in the sub-equilibrium regime in order to obtain reliable qualitative and quantitative predictions of the tribological response of layered materials interfaces under high external loads. Specifically, the fact that the ILP-MBD-bulk parameterization provides a better fit to the reference DFT data across the entire interlayer distance regime suggests that a linear friction dependence on pressure should be expected for this system.

## 6. Conclusion

In summary, we studied the reliability of HSE+TS and HSE+MBD DFT calculations for the description of the interlayer interactions in graphite and $h$-BN at sub-equilibrium interlayer distances. This was achieved by parameterizing our anisotropic ILP against the dispersion corrected DFT reference data, across a wide interlayer distance range. The ILPs were then used to perform fully atomistic MD simulations of bulk systems subjected to external pressure. By comparing the simulation results to experimental compressibility data of graphite and $h$-BN we found that the MBD-parameterized ILP provides better and satisfactory agreement with experiment up to pressures of 30 GPa and 14 GPa for graphite and $h$-BN, respectively. The bulk modulus of graphite, extracted from a Murnaghan plot obtained from the HSE+MBD parameterized ILP, was also found to be in good agreement with experimental data. Corresponding reference data for $h$-BN are too scattered. The agreement of calculated and experimental phonon spectra indicates that extending the applicability of our ILP to the deep sub-equilibrium interlayer distance regime does not sacrifice its ability to describe material properties at low external loads. Using the MBD parameterization, we were able to predict some structural and mechanical properties of two graphene/$h$-BN based heterostructures. We found that despite the highly corrugated superstructure formed, their load dependent interlayer distance is very similar to that of graphite and somewhat larger than in $h$-BN. The extracted bulk moduli of both heterogeneous structures were found to be comparable to those of the two homogeneous bulk systems investigated. Finally, dynamic friction simulations of the homogeneous systems suggest that the results strongly depend both qualitatively and quantitatively on the type of ILP parameterization. This demonstrates the importance of carefully choosing the DFT reference dataset for predicting the tribological properties of layered materials. The analysis performed in the present work, suggests that the ILP-MBD-bulk parameterization provides a better description of the interlayer interactions in homogeneous and heterogeneous junctions of graphene and $h$-BN, at a wide range of external loads.

We are currently extending the investigation to include transition-metal dichalcogenides such as $MoS_2$, $MoSe_2$, $WS_2$, and $WSe_2$. This will allow us to draw general conclusions regarding the applicability of the HSE+TS and HSE+MBD approximations and the corresponding ILPs for modeling layered materials subject to high external pressure.

ASSOCIATED CONTENT

**Supporting Information**.

The Supporting Information is organized as follows:

Interlayer potential (ILP) fitting parameters; Sensitivity test of the ILP parameters; Bulk modulus of graphite and hexagonal boron nitride; Sliding potential energy surfaces for bilayer configurations at sub-equilibrium interlayer distances; Dispersive component of the sliding energy surfaces; and Sliding energy barriers under different normal loads (PDF).

AUTHOR INFORMATION

**Corresponding Author**

* E-mail: odedhod@tauex.tau.ac.il.

**Notes**

The authors declare no competing financial interest.

## Acknowledgments

W.O. acknowledges the financial support from the Planning and Budgeting Committee fellowship program for outstanding postdoctoral researchers from China and India in Israeli Universities and the support from the National Natural Science Foundation of China (No. 11890673 & No. 11890674). M.U. acknowledges the financial support of the Israel Science Foundation, Grant No. 1141/18, and of the Deutsche Forschungsgemeinschaft (DFG), Grant No. BA 1008/21-1. O.H. is grateful for the generous financial support of the Israel Science Foundation under grant no. 1586/17 and the Naomi Foundation for generous financial support via the 2017 Kadar Award. This work is supported in part by COST Action MP1303. L.K. is the incumbent of the Aryeh and Mintzi Katzman Professorial Chair. D.M. acknowledges the fellowship from the Sackler Center for Computational Molecular and Materials Science at Tel Aviv University, and from Tel Aviv University Center for Nanoscience and Nanotechnology.

# Mechanical and Tribological Properties of Layered Materials Under High Pressure: Assessing the Importance of Many-Body Dispersion Effects Supporting Information

Wengen Ouyang,[1] Ido Azuri,[2] Davide Mandelli,[3] Alexandre Tkatchenko,[4,5] Leeor Kronik,[2]

Michael Urbakh,[1] and Oded Hod[1]

[1]*School of Chemistry and The Sackler Center for Computational Molecular and Materials Science, Tel Aviv University, Tel Aviv 6997801, Israel*

[2]*Department of Materials and Interfaces, Weizmann Institute of Science, Rehovoth 76100, Israel*

[3]*Istituto Italiano di Tecnologia, Via Morego, 30 16163 Genova, Italy*

[4]*Fritz-Haber-Institut der Max-Planck-Gesellschaft, Faradayweg 4-6, D-14195 Berlin, Germany*

[5]*Department of Physics and Materials Science, University of Luxembourg, Luxembourg*

This supporting information document includes the following sections:

1. Interlayer potential (ILP) fitting parameters.
2. Sensitivity test of the ILP parameters.
3. Bulk modulus of graphite and hexagonal boron nitride.
4. Sliding potential energy surfaces for bilayer configurations at sub-equilibrium interlayer distances.
5. Dispersive component of the sliding energy surfaces.
6. Sliding energy barriers under different normal loads.

# 1. Interlayer potential (ILP) fitting parameters

In this work, all reference data were obtained using dispersion-augmented density functional theory (DFT) calculations, which are based on the screened-exchange hybrid functional of Heyd, Scuseria, and Ernzerhof (HSE).[1-4] We employ both many-body dispersion (MBD)[5, 6] and Tkatchenko-Scheffler (TS) corrections[7, 8] to augment the HSE functional. In previous studies, the former scheme (HSE + MBD) was shown to provide a good balance between accuracy and computational burden for calculating binding energy curves and sliding energy landscapes for bilayer graphene, *h*-BN, and their heterojunctions.[9, 10] In recent work, we refined the ILP parameters to fit against the MBD corrected DFT reference for bilayer systems and improved the performance of the potential at the sub-equilibrium regime.[11] In the present study, to evaluate the properties of bulk materials, we performed DFT calculations for a fully periodic system (bulk configuration) with the same method. The resulting binding energy curves and sliding energy surfaces appear in Figures 1-3 of the main text. By using the fitting procedure introduced in ref [11], two sets of parameters, fitted against the HSE + MBD and HSE + TS DFT reference data, are given in **Table S1** and **Table S2**.

***Table S1.*** *List of ILP parameter values for bulk graphene and bulk h-BN based systems that are periodic in all directions. The training set includes all HSE + MBD binding energy curves and sliding potential surfaces appearing in Figs. 1-3 of the main text. A value of $R_{cut} = 16$ Å is used throughout.*

|     | $\beta_{ij}$ (Å) | $\alpha_{ij}$ | $\gamma_{ij}$ (Å) | $\varepsilon_{ij}$ (meV) | $C_{ij}$ (meV) | $d_{ij}$ | $s_{R,ij}$ | $r_{eff,ij}$ (Å) | $C_{6,ij}$ (eV·Å$^6$) | $\lambda_{ij}$ (Å$^{-1}$) |
|-----|--------|--------|--------|--------|--------|--------|--------|--------|--------|--------|
| C-C | 3.1894 | 8.2113 | 1.2600 | 0.0106 | 38.9821 | 10.9736 | 0.7869 | 3.4579 | 25.2496 | -- |
| B-B | 3.2147 | 7.1652 | 1.7459 | 11.0736 | 15.4819 | 15.4815 | 0.8550 | 3.4424 | 49.4984 | 0.70 |
| N-N | 3.3006 | 6.9226 | 1.4845 | 7.9908 | 46.6115 | 16.9081 | 0.7585 | 3.3266 | 14.8106 | 0.69 |
| B-N | 3.1709 | 8.5168 | 2.8657 | 5.4561 | 2.5548 | 13.5321 | 0.8863 | 3.4553 | 24.6708 | 0.694982 |
| C-B | 3.1007 | 5.1146 | 3.8387 | 18.2345 | 1.1902 | 10.2155 | 0.7686 | 3.5030 | 39.2629 | -- |
| C-N | 3.3173 | 10.3497 | 1.3793 | 16.3163 | 19.5691 | 15.7748 | 0.5645 | 3.2659 | 19.9631 | -- |

***Table S2.*** *List of ILP parameter values for bulk graphene and bulk h-BN based systems that are periodic in all directions. The training set includes all HSE + TS binding energy curves and sliding potential surfaces appearing in Figs. 1-3 of the main text. A value of $R_{cut} = 16$ Å is used throughout.*

|     | $\beta_{ij}$ (Å) | $\alpha_{ij}$ | $\gamma_{ij}$ (Å) | $\varepsilon_{ij}$ (meV) | $C_{ij}$ (meV) | $d_{ij}$ | $s_{R,ij}$ | $r_{eff,ij}$ (Å) | $C_{6,ij}$ (eV·Å$^6$) | $\lambda_{ij}$ (Å$^{-1}$) |
|-----|--------|--------|--------|--------|--------|--------|--------|--------|--------|--------|
| C-C | 3.1912 | 8.8423 | 1.1312 | 0.0863 | 33.4354 | 10.0196 | 0.9251 | 3.4842 | 32.4025 | -- |
| B-B | 3.5386 | 5.1268 | 2.2006 | 12.8753 | 27.5894 | 13.3600 | 0.8414 | 3.6431 | 99.5133 | 0.70 |
| N-N | 3.5915 | 3.2218 | 1.4354 | 6.6766 | 73.1026 | 13.0710 | 0.7466 | 3.3083 | 74.8236 | 0.69 |
| B-N | 3.9929 | 7.8553 | 2.5853 | 4.5785 | 2.3284 | 16.2665 | 0.8669 | 3.9824 | 84.7000 | 0.694982 |
| C-B | 3.0183 | 9.8126 | 3.6974 | 22.1591 | 0.8265 | 11.1783 | 0.9510 | 3.8465 | 40.1653 | -- |
| C-N | 3.4896 | 10.1614 | 1.1615 | 4.2615 | 11.1811 | 11.0391 | 0.9257 | 3.2512 | 29.0669 | -- |

## 2. Sensitivity test of the ILP parameters

In this section we investigate in some details the effects of the choice of the reference datasets (HSE + TS and HSE + MBD) on the ILP parametrization. For the sake of the discussion, we report here the analytical expression of the ILP:

$$V(r_{ij}, n_i, n_j) = \text{Tap}(r_{ij})[V_{\text{att}}(r_{ij}) + V_{\text{rep}}(r_{ij}, n_i, n_j) + V_{\text{Coul}}(r_{ij})],$$

where

$$\text{Tap}(r_{ij}) = 20\left(\frac{r_{ij}}{R_{\text{cut}}}\right)^7 - 70\left(\frac{r_{ij}}{R_{\text{cut}}}\right)^6 + 84\left(\frac{r_{ij}}{R_{\text{cut}}}\right)^5 - 35\left(\frac{r_{ij}}{R_{\text{cut}}}\right)^4 + 1$$

provides a continuous long-range cutoff (up to third derivative) that dampens the various interactions at interatomic separations larger than $R_{\text{cut}} = 16$ Å, and

$$V_{\text{Coul}}(r_{ij}) = kq_iq_j/\sqrt[3]{r_{ij}^3 + \lambda_{ij}^{-3}}$$

is the monopolar electrostatic interaction between atoms $i$ and $j$. We note that the parameters, $\lambda_{ij}$, and the atomic charges, $q_i$, are the same for both the HSE + TS and HSE + MBD parametrizations. Hence, to understand the effects of the chosen model on the ILP, we can consider only the terms $V_{\text{att}}$ and $V_{\text{rep}}$, corresponding to the long-range van der Waals attraction and short-range Pauli repulsion, respectively:

$$V_{\text{att}}(r_{ij}) = -\frac{1}{1 + e^{-d_{ij}[r_{ij}/(s_{R,ij} \cdot r_{ij}^{\text{eff}})-1]}} \frac{C_{6,ij}}{r_{ij}^6}$$

$$V_{\text{rep}}(r_{ij}, n_i, n_j) = e^{\alpha_{ij}\left(1-\frac{r_{ij}}{\beta_{ij}}\right)}\left\{\varepsilon_{ij} + C_{ij}\left[e^{-(\rho_{ij}/\gamma_{ij})^2} + e^{-(\rho_{ji}/\gamma_{ij})^2}\right]\right\}$$

Here, $r_{ij}$ is the Euclidean distance between the two atoms involved, $n_i$ is the surface normal at the position of atom $i$ and $\rho_{ij}$ is the lateral distance between the normal vectors at the positions of atoms $i$ and $j$.

To study the effects of the chosen model on the ILP, in the first row of **Figure S1** we compare the ILP-TS-bulk and ILP-MBD-bulk binding energy curves computed for three periodic systems: graphite, bulk $h$-BN and alternating graphene/$h$-BN. For all cases considered, the minimum of the ILP-TS-bulk curve (dashed blue lines in **Figure S1**) is $\lesssim 50$ meV/atom lower than the value predicted by the ILP-MBD-bulk parametrization (continuous red lines). This is accompanied by only minor changes in the equilibrium interlayer distance (differences $\lesssim 3\%$, as reported also in Table 1 and Table 2 of the main text). Differences between the HSE + TS and HSE + MBD parameterized ILP curves become negligible at interlayer distances $d \gtrsim 6$ Å.

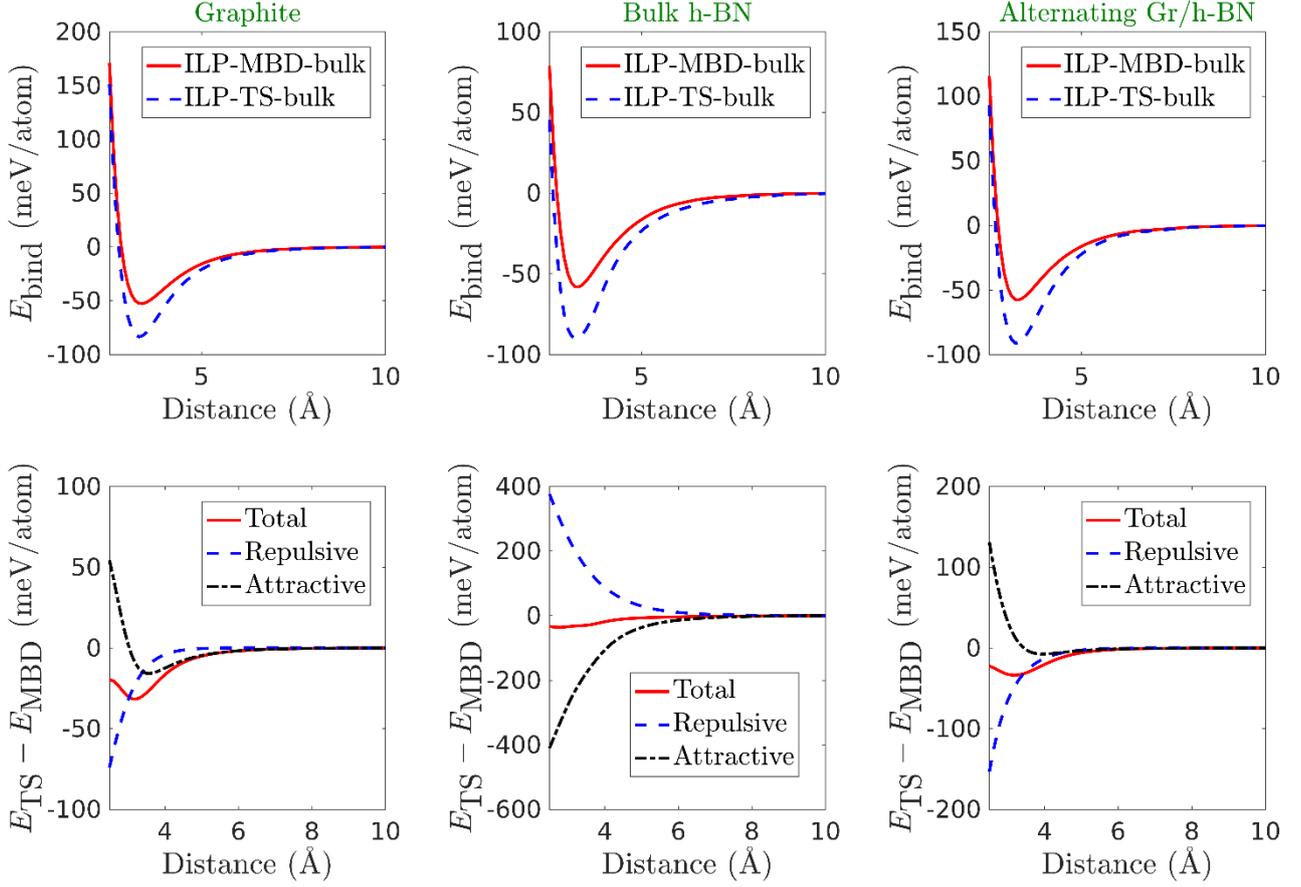

***Figure S1.*** *Comparison between HSE + TS and HSE + MBD parametrized ILP binding energy curves for graphite (left column), bulk h-BN (middle column), and bulk alternating graphene/h-BN. The first row reports the comparison between the binding energy curve corresponding to the MBD (solid red line) and TS (blue dashed line) parametrization of the ILP, for three systems: graphite (left panel), bulk h-BN (middle panel), and alternating graphene/h-BN (right panel). The second row reports the corresponding differences between the ILP-TS-bulk and ILP-MBD-bulk binding energy curves (solid red lines), between the MBD and TS Pauli repulsion components (blue dashed line) and between the ILP-MBD-bulk and ILP-TS-bulk van der Waals attractive components (dotted-dash black line).*

To better understand the origin of the observed variations, in the second row of **Figure S1** we report the difference between the ILP-MBD-bulk and ILP-TS-bulk binding energy curves (red continuous lines), together with the difference computed considering only the repulsive (dashed blue lines) or the attractive (dash-dotted black lines) terms. For the case of graphite, the TS parametrization predicts larger attraction at interlayer distances $d \lesssim 3$ Å, which becomes smaller than the MBD prediction between $3 \lesssim d \lesssim 6$ Å. A detailed analysis of the effects of each single parameter on the ILP reveals that these two outcomes are due to the changes of the $C_{6,ij}$ and $s_{R,ij}$ parameters, respectively (see first row of **Table S3**, and last row of **Figure S2**). The Pauli repulsion predicted by the TS parametrization is instead always smaller than the one predicted the MBD parametrization. This is mainly caused by the variation of the $C_{ij}$ and $\gamma_{ij}$ parameters (see first two rows of **Figure S2**). For the case of bulk h-BN (middle panels in **Figure S1**), the repulsive and attractive parts computed via the TS parametrization are respectively larger and smaller than the corresponding MBD values. The origin of these differences are mainly due to the changes of the $\beta_{ij}$, $C_{ij}$, and $\alpha_{ij}$ parameters for the repulsive part, and to the changes of the $C_{6,ij}$ parameter for the attractive part (see **Table S3** and **Figure S3**). Finally, for the case of the alternating graphene/h-BN system, we observe an opposite behavior, where the repulsive and attractive interactions predicted by the TS parametrization are respectively smaller and larger than the corresponding MBD values (see bottom right panel in **Figure S1**). The detailed analysis reported in **Figure S4** demonstrates that this behavior arises from the interplay of several parameters.

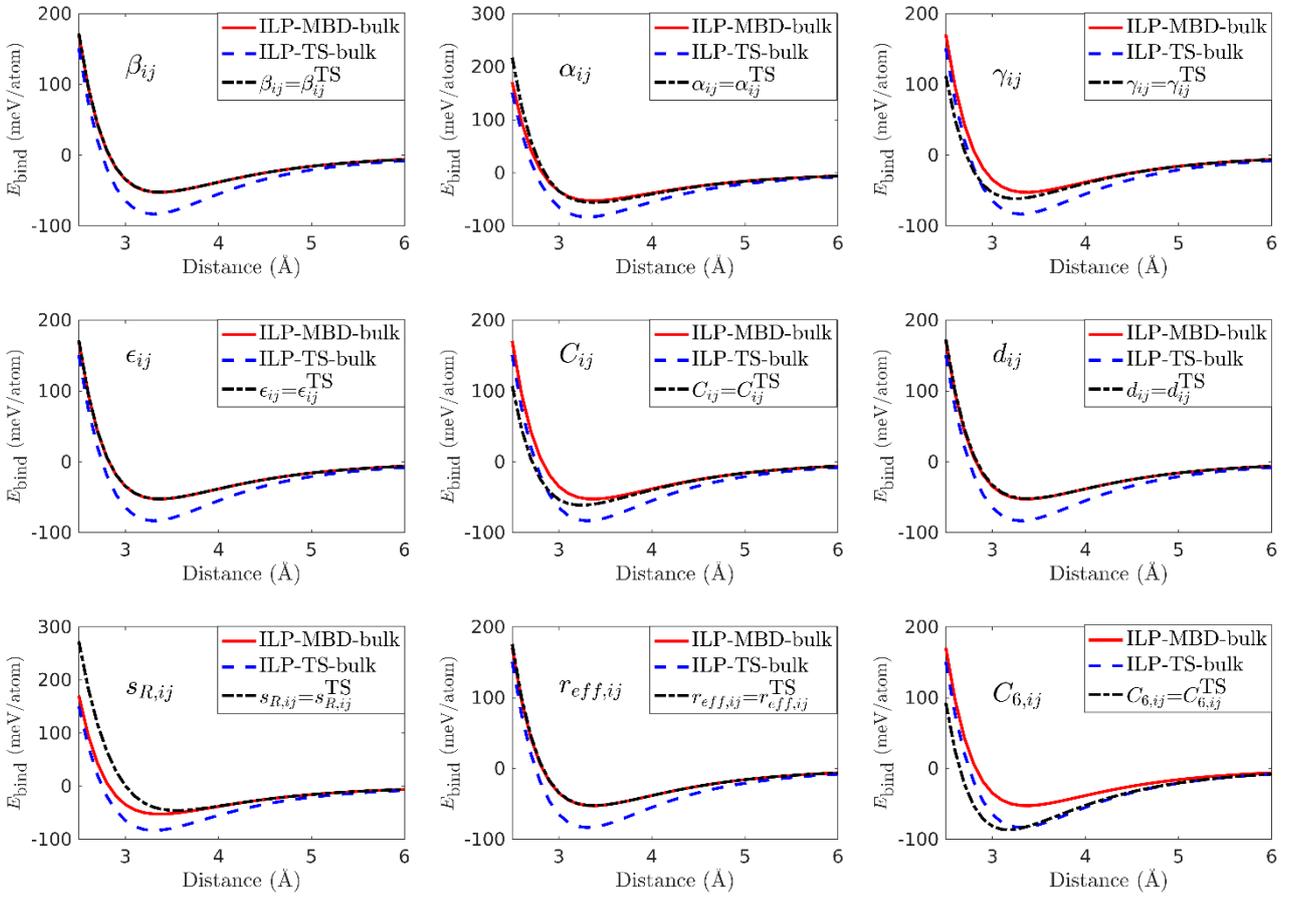

***Figure S2.*** *Sensitivity of the ILP to changes in parameter values for bulk graphite. In each panel, the red solid line and blue dashed line are binding energy curves computed using the MBD and TS parameterizations, respectively. The black dash-dotted line is the binding energy curve calculated using the MBD values for all parameters except one (labeled in each panel), which is changed to the corresponding TS values, for each distinct pair of atoms, as reported in **Table S3**.*

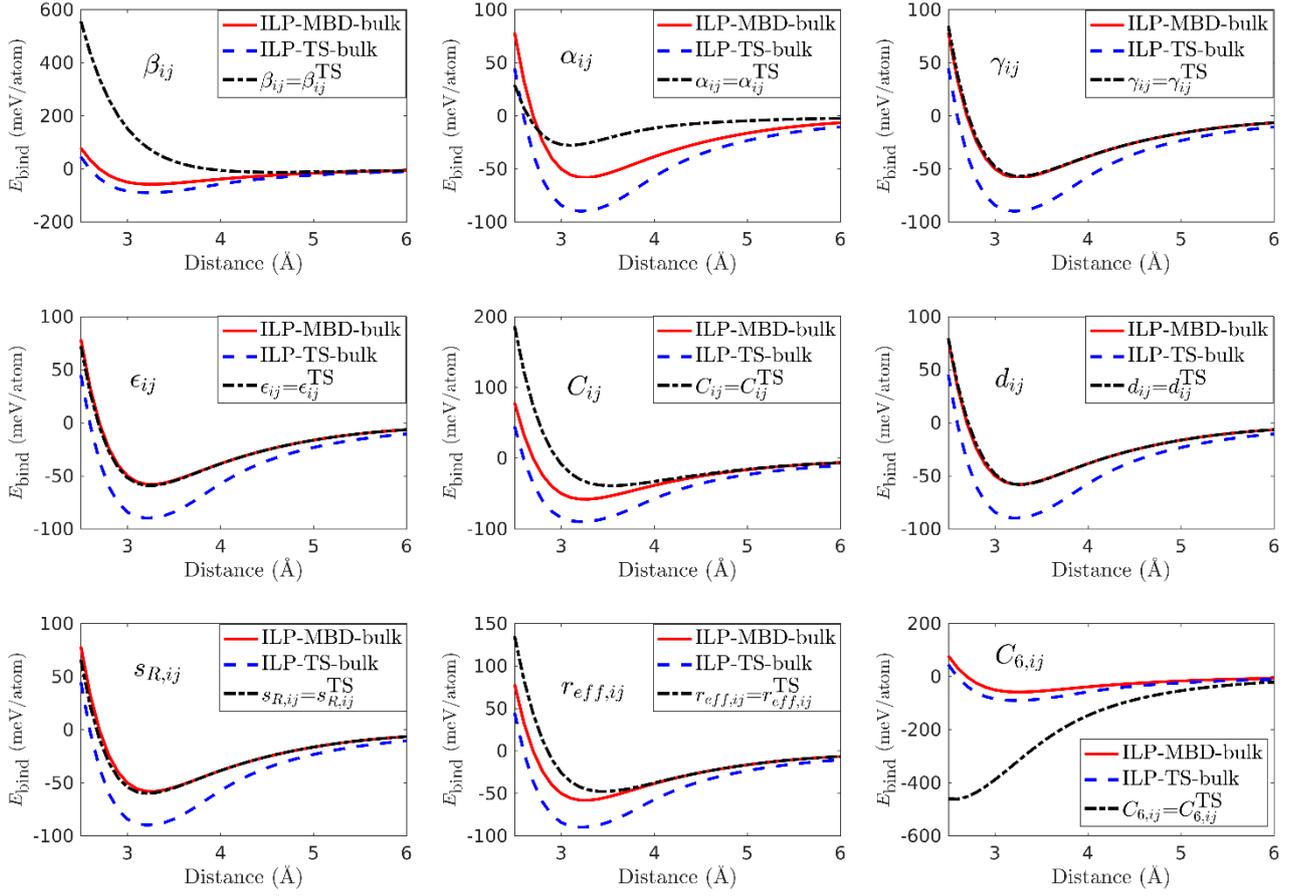

*Figure S3.* Sensitivity of the ILP to changes in the parameter values for bulk h-BN. In each panel, the red solid line and blue dashed line are binding energy curves computed using the MBD and TS parameterizations, respectively. The black dash-dotted line is the binding energy curve calculated using the MBD values for all parameters except one (labeled in each panel), which is changed to the corresponding TS values, for each distinct pair of atoms, as reported in *Table S3*.

*Table S3.* For each pair of atom, we report the change, $\Delta X = (X_{TS} - X_{MBD})/X_{MBD}$, of the various ILP parameters obtained from the HSE + TS parametrization, relative to the value obtained from the HSE + MBD parametrization.

|  | $\Delta\beta_{ij}$ (%) | $\Delta\alpha_{ij}$ (%) | $\Delta\gamma_{ij}$ (%) | $\Delta\varepsilon_{ij}$ (%) | $\Delta C_{ij}$ (%) | $\Delta d_{ij}$ (%) | $\Delta s_{R,ij}$ (%) | $\Delta r_{eff,ij}$ (%) | $\Delta C_{6,ij}$ (%) | $\Delta\lambda_{ij}$ (%) |
|---|---|---|---|---|---|---|---|---|---|---|
| **C-C** | 0.06 | 7.7 | -10 | 714 | -14 | -8.8 | 18 | 0.76 | 28 | -- |
| **B-B** | 10.0 | -28 | 26 | 16 | 78 | -14 | -1.6 | 5.8 | 101 | 0 |
| **N-N** | 8.8 | -53 | -3.3 | -16 | 57 | -23 | -1.6 | -0.55 | 405 | 0 |
| **B-N** | 26 | -7.8 | -9.8 | -16 | -8.9 | 20 | -2.2 | 15 | 243 | 0 |
| **C-B** | -2.7 | 92 | -3.7 | 22 | -31 | 9.4 | 24 | 9.8 | 2.3 | -- |
| **C-N** | 5.2 | -1.8 | -16 | -74 | -43 | -30 | 64 | -0.45 | 46 | -- |

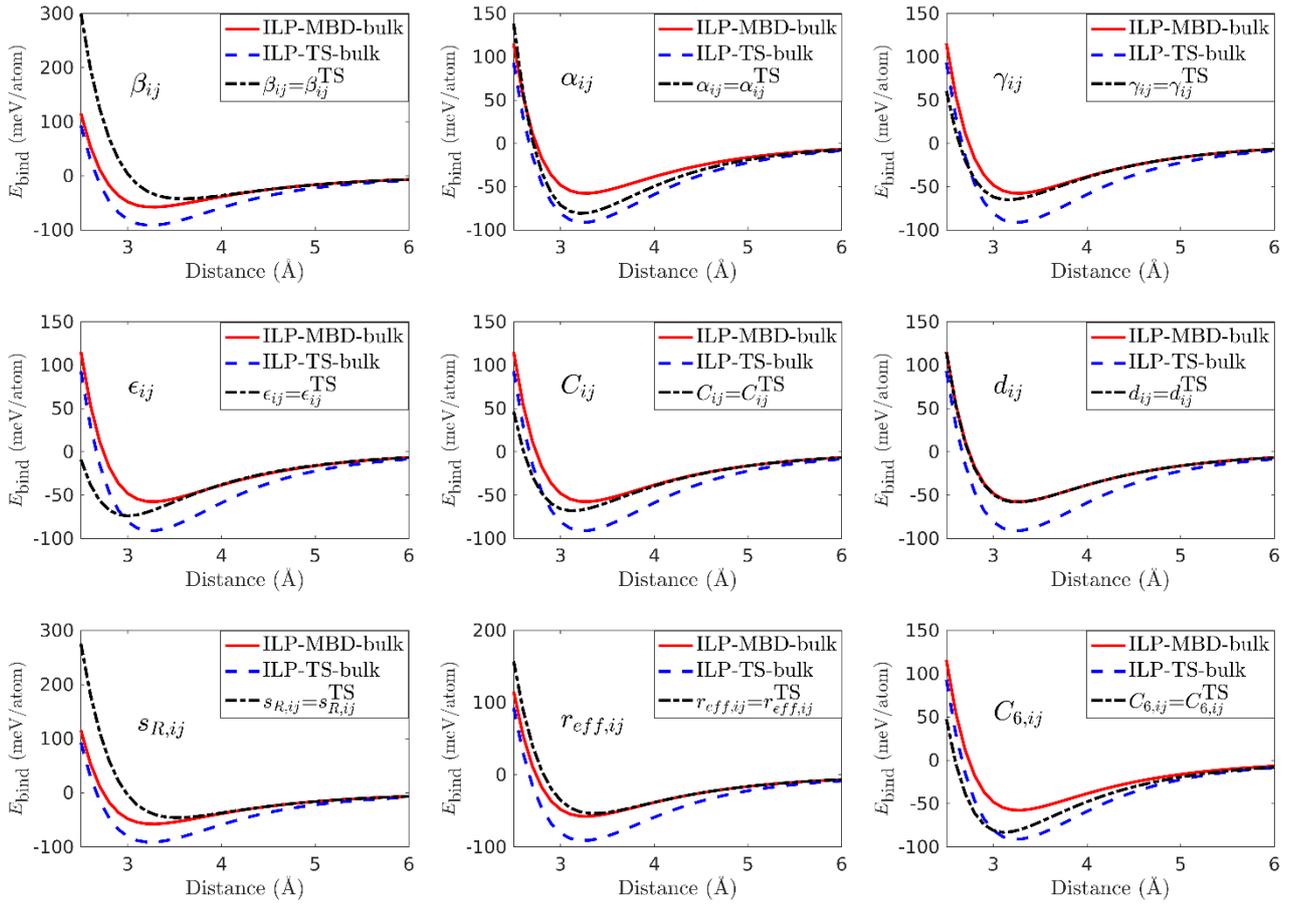

*Figure S4. Sensitivity of the ILP to changes in parameter values for bulk alternating graphene/h-BN configuration. In each panel, the red solid line and blue dashed line are binding energy curves computed using the MBD and TS parameterizations, respectively. The black dash-dotted line is the binding energy curve calculated using the MBD values for all parameters except one (labeled in each panel), which is changed to the corresponding TS values, for each distinct pair of atoms, as reported in **Table S3**.*

Overall, the above analysis that focused on binding energy curves suggests that changing the reference model affects different parameters in different ways, depending on the material considered. The combined effects of such changes on the ILP determines the final shape of the binding energy curves. This, of course, is a general feature of force-field parameterizations. While all parameter values are kept within reasonable physical ranges during the optimization procedure, discussing separately the specific value of each parameter goes beyond the accuracy limits of the method and only their combined behavior should be considered. Nevertheless, from a careful inspection of **Figures S2**-**S4** it becomes clearly evident that the binding energy curve can be very sensitive to the value of some parameters, especially the isotropic long-range attraction $C_{6,ij}$ coefficients and the anisotropic repulsion $C_{ij}$ coefficients. Therefore, extra care should be taken when fitting their values.

## 3. Bulk modulus of graphite and hexagonal boron nitride

**Figure S5** shows the normalized volume $V/V_0$ ($V_0$ being the volume at zero pressure) of bulk graphite and bulk *h*-BN, as a function of pressure. The open symbols represent equilibrium molecular dynamics (EMD) simulation results obtained with different ILP and Kolmogorov-Crespi (KC) potential parameterizations.[9-12] The solid lines are the fitted Murnaghan equation (eq 1 in the main text) results.[13] The fitted parameters (bulk modulus and its pressure derivative) are listed in Tables 1-2 in the main text for bulk graphite and bulk *h*-BN, respectively.

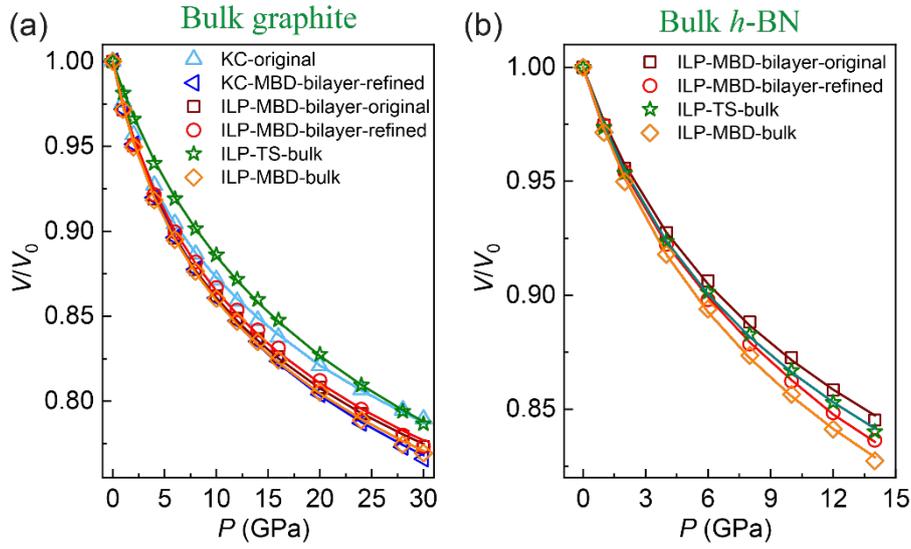

*Figure S5. Pressure dependence of the normalized volume $V/V_0$ of bulk graphite and bulk h-BN. The open points are the NPT simulations results for different parameterizations of the ILP and KC potentials. The solid lines are fitted curves generated by eq 1 in the main text.*

It should be noted that apart from the Murnaghan equation, two other equations of state (EOS) are also commonly used to fit the *P-V* curve: (i) The Birch-Murnaghan equation (eq S1)[14, 15] and (ii) The Vinet equation (eq S2),[16, 17] which take the following forms:

$$P = 3B_V^0 \xi (1+2\xi)^{5/2} \left[1 - \frac{3}{2}(4-B_V')\xi\right], \quad \xi = \frac{1}{2}\left[\left(\frac{V}{V_0}\right)^{-\frac{2}{3}} - 1\right], \tag{S1}$$

$$P = 3B_V^0 \frac{(1-X)}{X^2} \exp\left[\frac{3}{2}(B_V'-1)(1-X)\right], \quad X = \left(\frac{V}{V_0}\right)^{\frac{1}{3}}. \tag{S2}$$

As in the Murnaghan equation, these two EOS also assume that $B_V$ varies with pressure (hence the inclusion of $B_V'$). Nonetheless, they differ in their description of the dependence of $B_V$ on the pressure, by assuming that it is linear, polynomial, and exponential for the Murnaghan, Birch–Murnaghan, and Vinet EOS, respectively.

**Table S4** lists the fitting EMD results for the bulk modulus of graphite and bulk *h*-BN with the three commonly used EOS. Unlike the observations from a previous study,[18] where the bulk modulus was found to be very sensitive to the choice of EOS, here we find that all three EOS yield comparable values for the bulk modulus. This suggests that the differences between the experimental values of the bulk modulus arise from the different measuring techniques adopted in different studies rather than from the choice of the EOS used for the fitting procedure.

*Table S4. Bulk moduli obtained by fitting our EMD data with different equations of state for graphite and bulk h-BN. Experimental values are presented for comparison.*

| Material | Methods | Murnaghan | | Birch-Murnaghan | | Vinet | |
|---|---|---|---|---|---|---|---|
| | | $B_V^0$ (GPa) | $B_V'$ | $B_V^0$ (GPa) | $B_V'$ | $B_V^0$ (GPa) | $B_V'$ |
| Graphite | Experiments | 33±2[a] | 12.3±0.7[a] | -- | -- | -- | -- |
| | | 33.8±0.3[b] | 8.9±0.1[b] | -- | -- | -- | -- |
| | | -- | -- | -- | -- | 30.8±2[c] | -- |
| | ILP-MBD-bulk | 34±1 | 8.1±0.3 | 27±1 | 14.2±0.7 | 31.5±0.8 | 10.2±0.2 |
| | ILP-TS-bulk | 55±2 | 6.2±0.3 | 53±0.9 | 7.5±0.2 | 53.4±0.9 | 7.4±0.2 |
| | ILP-MBD-bilayer-refined, ref 11 | 36±3 | 8.1±0.6 | 33±2 | 12.2±0.9 | 36±2 | 9.6±0.5 |
| | ILP-MBD-bilayer-original, ref 10 | 33±1 | 8.5±0.3 | 25.5±0.8 | 16.3±0.7 | 30.7±0.3 | 10.8±0.1 |
| | KC-MBD-bilayer-refined, ref 11 | 35±2 | 7.7±0.3 | 30.5±0.5 | 12.0±0.3 | 33.5±0.7 | 9.5±0.2 |
| | KC-original, ref 12 | 37±2 | 8.9±0.4 | 29.3±0.6 | 16.7±0.4 | 35.1±0.7 | 11.1±0.2 |
| Bulk *h*-BN | Experiments | 22±4[a] | 18±3[a] | -- | -- | -- | -- |
| | | 36.7±0.5[d] | 5.6±0.2[d] | -- | -- | -- | -- |
| | | -- | -- | 17.6±0.8[e] | 19.5±3.4[e] | -- | -- |
| | | -- | -- | 27.6±0.5[f] | 10.5±0.5[f] | -- | -- |
| | ILP-MBD-bulk | 33±2 | 7.8±0.6 | 31±1 | 10.2±0.8 | 32±1 | 9.0±0.5 |
| | ILP-TS-bulk | 35±2 | 8.7±0.6 | 33±1 | 12.0±0.7 | 34±1 | 10.0±0.5 |
| | ILP-MBD-bilayer-refined, ref 11 | 35±2 | 8.0±0.6 | 33±1 | 10.5±0.7 | 34±1 | 9.2±0.5 |
| | ILP-MBD-bilayer-original, ref 10 | 38±3 | 8.7±0.9 | 36±2 | 11±1 | 38±2 | 9.7±0.9 |

[a]ref 19,   [b]ref 20,   [c]ref 21,   [d]ref 22,   [e]ref 23,   [f]ref 24.

# 4. Sliding potential energy surfaces for bilayer configurations at sub-equilibrium interlayer distances

Because the repulsive walls of the binding energy curves are very steep at the sub-equilibrium interlayer distance regime, the differences between energy and forces calculated using different methods are expected to increase in absolute value in this range. To demonstrate this, we present in **Figure S6-Figure S8** the sliding potential energy surfaces for periodic bilayer graphene and bilayer *h*-BN, calculated using the refined ILP and KC potential[11] as well as the original ILP[10] and KC potentials,[12] at three sub-equilibrium interlayer distances. The first and second rows in **Figure S6** present the sliding energy surfaces of periodic bilayer graphene with interlayer distances of 3.35 Å (left column), 3.0 Å (middle column) and 2.8 Å (right column), calculated using the refined[11] and original ILP,[10] respectively. The differences between the two are presented in the third row of the figure. Clearly, the differences between the sliding energy surfaces obtained using the two parameterizations increase in both magnitude and relative value as the interlayer distance decreases. Specifically, the maximal absolute differences obtained are 0.4 (~2%), 2 (~13%), and 8 meV/atom (~20%) for interlayer distances of 3.35 Å, 3.0 Å, and 2.8 Å, respectively.

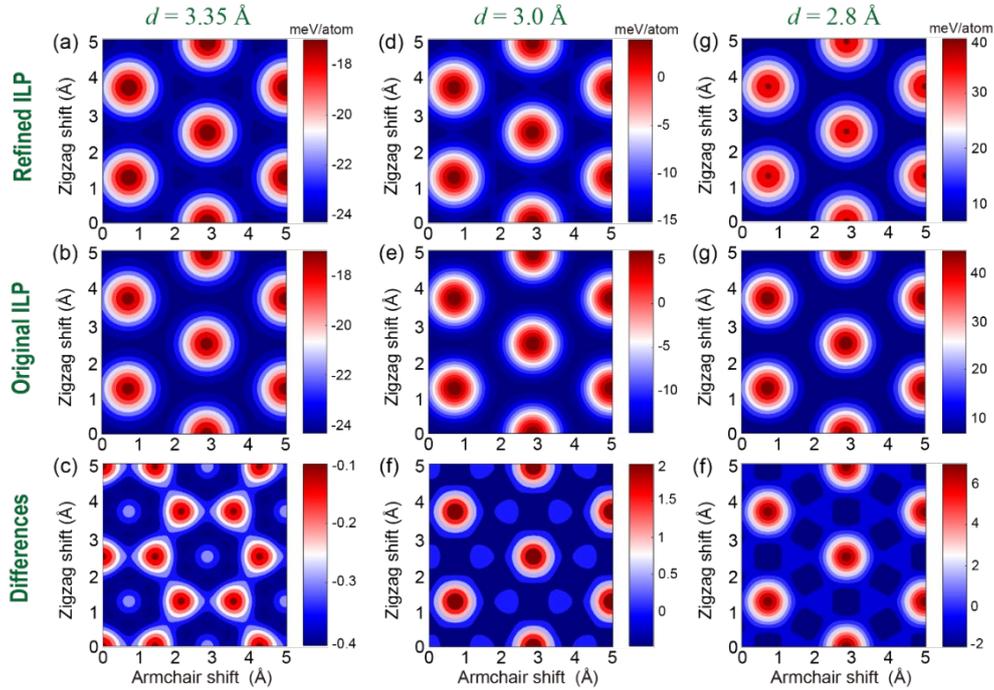

*Figure S6.* Sliding energy surfaces of periodic bilayer graphene for three different interlayer distances. The first and second rows present the sliding energy surfaces obtained at interlayer distances of 3.35 Å (left column), 3.0 Å (middle column), and 2.8 Å (right column) calculated using the refined [11] and original graphene ILP,[10] respectively. The third row presents the differences between the results obtained using the two ILP parameterizations.

The first and second rows in **Figure S7** present the sliding energy surfaces of periodic bilayer graphene with interlayer distances of 3.35 Å (left column), 3.0 Å (middle column) and 2.8 Å (right column) calculated using the refined and original KC potential, respectively. The differences between the two are presented in the third row of the figure. Clearly, the differences between the sliding energy surfaces obtained using the two parameterizations increase in both magnitude and relative value as the interlayer distance decreases. Specifically, the maximal absolute differences obtained are 2.2 (~10%), 5.3 (~20%), and 16.4 meV/atom (~40%), for interlayer distances of 3.35 Å, 3.0 Å, and 2.8 Å, respectively.

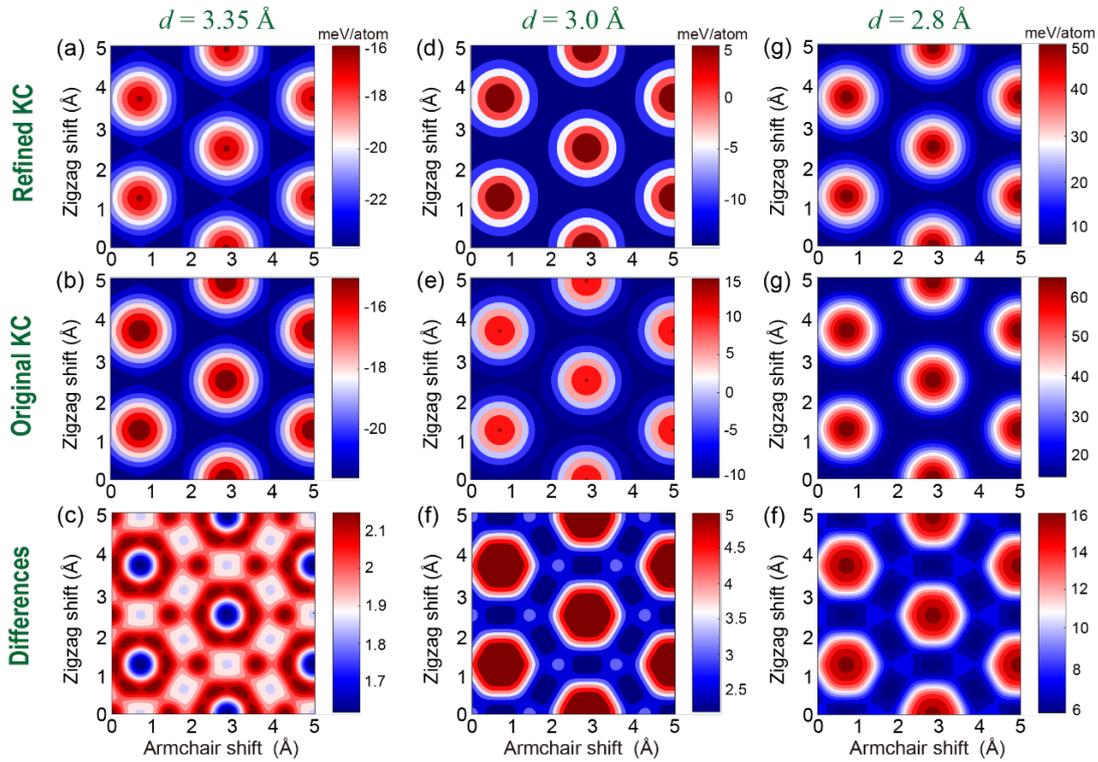

*Figure S7.* Sliding energy surfaces of periodic bilayer graphene for three different interlayer distances. The first and second rows present the sliding energy surfaces obtained at interlayer distances of 3.35 Å (left column), 3.0 Å (middle column), and 2.8 Å (right column) calculated using the refined [11] and original KC,[12] respectively. The third row presents the differences between the results obtained using the two KC parameterizations.

The first and second rows in **Figure S8** present the sliding energy surfaces of periodic bilayer $h$-BN with interlayer distances of 3.3 Å (left column), 3.0 Å (middle column), and 2.8 Å (right column) calculated using the refined and original ILP, respectively. The differences between the two are presented in the third row of the figure. Clearly, the differences between the sliding energy surfaces obtained using the two parameterizations increase in both magnitude and relative value as the interlayer distance decreases.

Specifically, the maximal absolute differences obtained are 0.49 (~4%), 1.8 (~10%), and 4.3 meV/atom (~40%), for interlayer distances of 3.3 Å, 3.0 Å, and 2.8 Å, respectively.

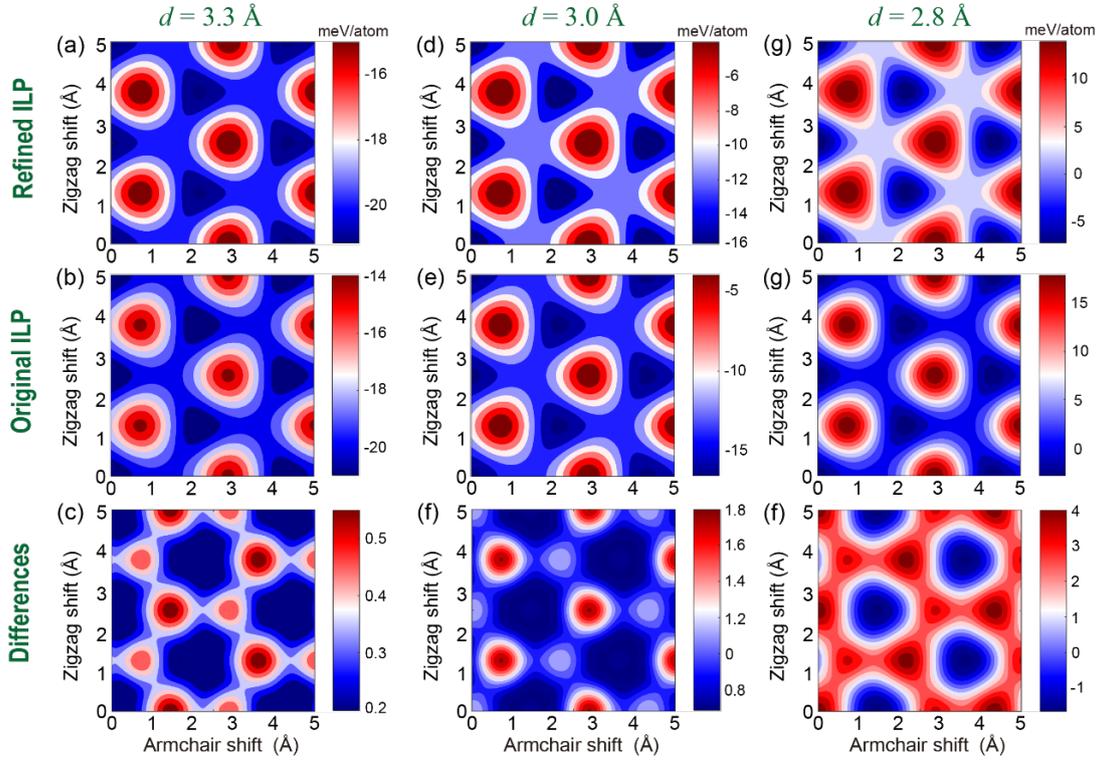

*Figure S8. Sliding energy surfaces of periodic bilayer h-BN for three different interlayer distances. The first and second rows present the sliding energy surfaces obtained at interlayer distances of 3.3 Å (left column), 3.0 Å (middle column), and 2.8 Å (right column) calculated using the refined [11] and original h-BN ILP,[10] respectively. The third row presents the differences between the results obtained using the two ILP parameterizations.*

## 5. Dispersive component of the sliding energy surfaces.

To evaluate the ability of the ILP to capture the dispersive component contribution to the sliding energy surface corrugation we plot in the second row of **Figure S9** the differences between the HSE + MBD results and the HSE only results (see first row of the figure) for the bulk graphite (left column), bulk *h*-BN (middle column), and alternating graphene/*h*-BN (left column) systems. Similar results for the TS dispersive component appear in the third row of the figure. Both the MBD and TS dispersive components are found to be typically lower than 2 meV/atom (apart from the TS component of the *h*-BN system that shows a corrugation of ~4 meV/atom), which is below the accuracy of the ILP fitting to the full HSE + MBD and HSE + TS reference data for these systems (see lower rows of Figures 2 and 3 of the main text). This indicates that the ILP cannot be expected to capture the dispersive component contribution to the sliding energy surface alone for the systems considered. We note in passing that, while the HSE contribution (first row of **Figure S9**) does not quantitatively capture the sliding energy surface, it is able to capture its overall symmetry obtained by the dispersion augmented methods.

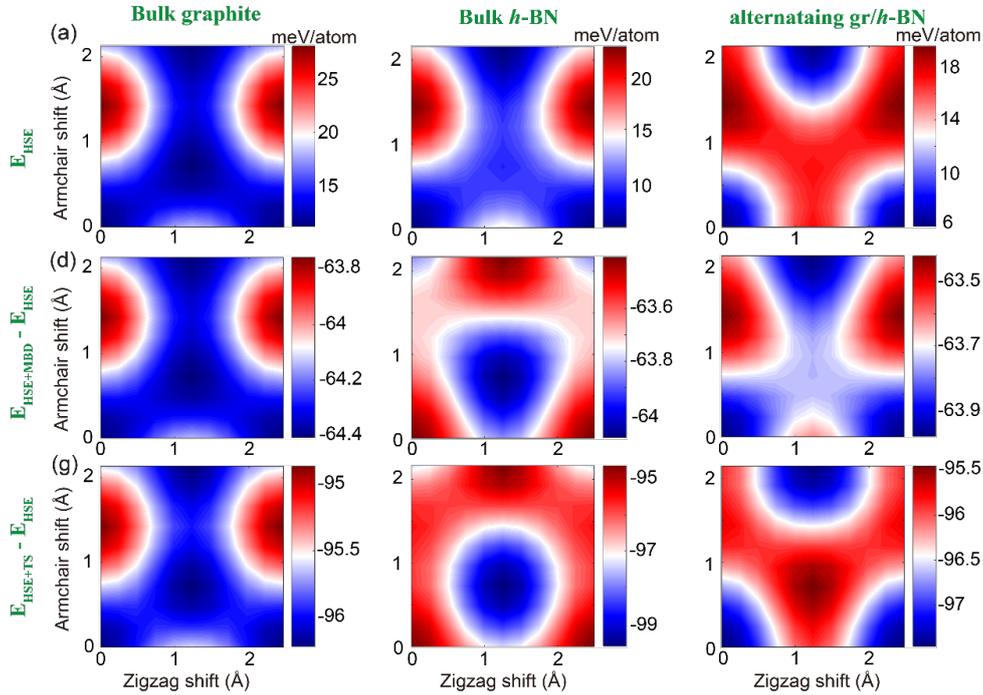

*Figure S9. Dispersive component contribution to the sliding energy surfaces of the periodic structures considered, calculated at an interlayer distance of 3.3 Å. The first row presents the sliding energy surface of bulk graphite (left panels), bulk h-BN (middle panels), and an alternating graphene/h-BN stack (right panels) systems, calculated using HSE. The MBD and TS dispersion contributions to the sliding energy surfaces are presented in the second and third rows, respectively. These are obtained by subtracting the HSE surface from either the HSE + MBD or the HSE + TS results. The reported energies are measured relative to the value obtained for the infinitely separated layers and are normalized by the total number of atoms in the unit-cell.*

## 6. Sliding energy barriers under different normal loads

To rationalize the differences in the friction forces obtained using the different ILP parameterizations (see Fig. 8 of the main text), we plot the energy barriers encountered during the sliding process as a function of the applied normal load for the four-layer graphene (**Figure S10**a) and *h*-BN (**Figure S10**b) model systems. For each stick-slip event, the energy barrier is evaluated from the ILP energy difference between the pre-slip and post-slip states. **Figure S10** presents the overall energy barrier, $U_{sl}$, obtained by averaging the results over several stick-slip events during steady-state sliding. The error bars represent the corresponding standard deviation resulting from thermal fluctuations. As can be seen, the friction force dependence on the normal load, presented in Figure 8 of the main text, follows the trends exhibited by the sliding energy barriers for the different ILP parameterizations.

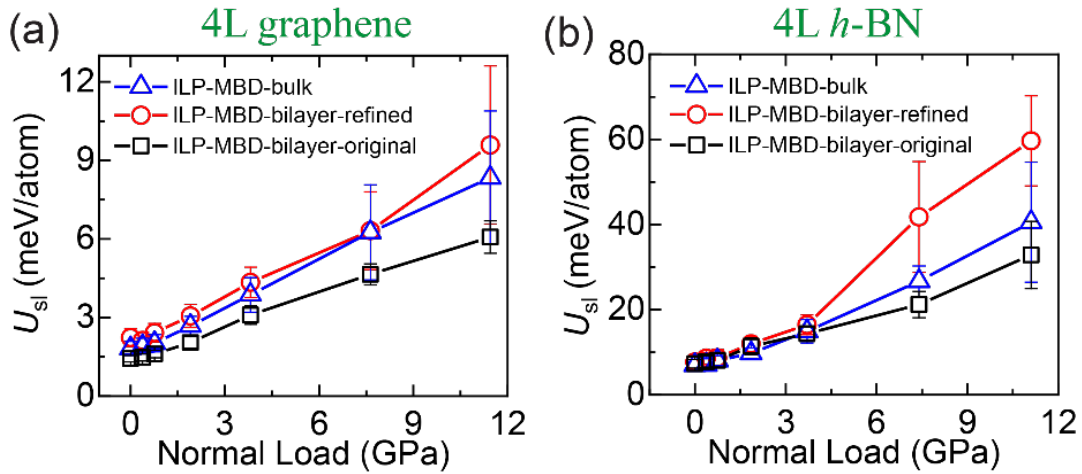

*Figure S10. Normal load dependence of the sliding potential energy barriers obtained for model systems consisting of four layers of (a) graphene and (b) h-BN. The simulations are performed at a temperature of 300 K for three different ILP parameterizations as listed in the figure.*